\documentclass{pas}
\usepackage{url}
\usepackage{hyperref}

\usepackage{multirow}
\usepackage{natbib}
\usepackage{graphicx}
\usepackage{subcaption}
\usepackage{aas_macros}
\begin{document}

\lefttitle{Reliable Polarimetry with uGMRT Band 4}
\righttitle{Pal et al.}

\jnlPage{1}{20}
\jnlDoiYr{2026}
\doival{10.1017/pasa.xxxx.xx}

\articletitt{Research Paper}

\title{Reliability of uGMRT Band-4 Polarimetry: Results from a Quadrature Hybrid polariser Bypass Experiment}

\author{\sn{Arpan} \gn{Pal}$^{1,2}$ and \sn{Sanjeet} \gn{Rai}$^{3}$ and \sn{Ganesh} \gn{Kumbhar}$^{3}$}

\affil{$^1$National Centre for Radio Astrophysics, Tata Institute of Fundamental Research, S. P. Pune University Campus, Ganeshkhind, Pune, 411007, India}
\affil{$^2$ National Radio Astronomy Observatory, P.O. Box O, Socorro, NM 87801, USA}
\affil{$^3$Giant Metrewave Radio Telescope, NCRA-TIFR, Khodad, Pune, 410504, India}

\corresp{A. Pal, Email: arpan522000@gmail.com}

\citeauth{Pal A, Rai S and Kumbhar G, Reliability of uGMRT Band-4 Polarimetry: Results from a Quadrature Hybrid polariser Bypass Experiment. {\it Publications of the Astronomical Society of Australia} {\bf 00}, 1--20}

\history{(Received xx xx xxxx; revised xx xx xxxx; accepted xx xx xxxx)}

\begin{abstract}

Polarimetric observations at sub-GHz frequencies offer unique access to the magnetized universe through Faraday rotation and depolarization studies, but achieving reliable polarization calibration at these frequencies remains challenging. We report the identification and resolution of a systematic polarization calibration instability in the upgraded Giant Metrewave Radio Telescope (uGMRT) Band 4 (550--750\,MHz). Through diagnostic observations of multiple calibrators, we discovered that the cross-hand phase response varies with the fractional polarization of the observed source, violating the fundamental assumption of calibration transferability in radio interferometry. Systematic engineering tests traced this behaviour to the Quadrature Hybrid (QH) polariser in the frontend signal chain. We conducted a controlled experiment in which the QH was bypassed in seven antennas, converting them to linear polarization feeds. The bypassed system shows dramatically improved performance: instrumental leakage reduced from 10--15\% to 2--5\%, residual leakage after calibration reduced from $\sim$0.5\% to less than $0.2\%$, and stable cross-hand phases independent of source polarization. For the polarized source DA\,240 (RM\,$=$\,3.3\,rad\,m$^{-2}$), the QH-bypassed system accurately recovers the expected $25^\circ$ polarization angle rotation across the band, which the with QH system fails to reproduce. These results establish that the QH polariser is the dominant source of polarimetric instability in uGMRT Band\,4 and demonstrate that its removal enables reliable sub-GHz polarimetry. We recommend the linear feed configuration for science cases requiring accurate polarization angle and rotation measure measurements.

\end{abstract}


\maketitle

\section{Introduction}

Radio telescopes measure the properties of the incoming electromagnetic radiation using feeds sensitive to two orthogonal polarization states. This makes radio interferometers inherently suitable for polarimetric observations, by measuring the full Stokes parameters ($I$, $Q$, $U$, $V$) that completely characterize the polarization state of the radiation \citep{1986isra.book.....T}.

The response of a radio telescope to incoming radiation can be elegantly described using the Jones matrix formalism \citep{1996A&AS..117..137H, 1996A&AS..117..149S,2011A&A...527A.106S}. The electric field of an incoming electromagnetic wave can be represented as a two-component complex vector $\mathbf{e}$. The measured voltage $\mathbf{v}$ at the output of a feed is related to the true sky electric field through a $2 \times 2$ complex Jones matrix $\mathbf{J}$:
\begin{equation}
\mathbf{v} = \mathbf{J} \mathbf{e}
\end{equation}
The total Jones matrix of an antenna can be decomposed into a product of matrices, each representing a distinct instrumental or propagation effect. For an interferometer baseline formed between antennas $i$ and $j$, the measured visibility matrix $\mathbf{V}_{ij}^{\mathrm{obs}}$ is related to the true sky coherency matrix $\mathbf{V}_{ij}^{\mathrm{sky}}$ by:
\begin{equation}
\mathbf{V}_{ij}^{\mathrm{obs}} = \mathbf{J}_i \mathbf{V}_{ij}^{\mathrm{sky}} \mathbf{J}_j^{\dagger}
\end{equation}
where the dagger denotes the conjugate transpose.

The calibration of a radio interferometer involves solving these Jones matrices that describe the instrumental response by observing appropriate calibrator sources. The standard calibration chain includes residual delay calibration, time and frequency dependent gain corrections using known calibrator sources. These steps primarily address the parallel-hand correlations and assume that the Jones matrix is approximately diagonal, which is generally sufficient for total intensity observations. For full polarization calibration, additional steps are required to determine the off-diagonal elements of the Jones matrix. These steps include estimating the instrumental leakage terms by observing an unpolarized source, or a source with known polarization properties, or a polarized source over a wide range of parallactic angles \citep{1996A&AS..117..137H,1996A&AS..117..149S,2021hai1.book..127R}. Another step corrects the differential delay between the two polarization channels, which introduces a linear phase slope across frequency in the cross-hand correlations \citep{1996A&AS..117..149S,2021hai1.book..127R}. This delay is measured using a polarized calibrator. The final step calibrates the absolute polarization angle using a source with a well-determined polarization orientation \citep{1996A&AS..117..149S,2021hai1.book..127R}. This final step is crucial for circular feeds, which do not have a fixed absolute EVPA reference on the sky. For linear feeds, however, the absolute EVPA can be set using the known orientation of the reference antenna’s feed. The calibration procedure relies on the principle that the instrumental response remains stable between calibrator and target observations or varies smoothly enough to be reliably interpolated.

The upgraded Giant Metrewave Radio Telescope (uGMRT) is a 30-element interferometer array with 45-meter dishes, operating across four frequency bands from 120 MHz to 1460 MHz \citep{2017CSci..113..707G}. Band 4 of the uGMRT covers from 550 to 950 MHz. uGMRT is the most sensitive radio interferometer in the Northern hemisphere to be operated at these frequencies. Sub-GHz polarimetry provides unique insights into Faraday rotation \citep{1966MNRAS.133...67B, 2005A&A...441.1217B}, depolarization mechanisms \citep{1998MNRAS.299..189S}, and the large-scale magnetic Universe \citep{2015A&ARv..24....4B}. Nevertheless, achieving reliable polarization calibration at these frequencies has long been a significant challenge.

During regular polarimetric observations with uGMRT Band 4, we encountered a systematic issue that hindered accurate polarization calibration. The calibrated data exhibited persistent phase ramps in the cross-hand visibilities, which could not be corrected using standard polarization calibration procedures, including leakage correction, cross-hand delay estimation, and polarization angle calibration. The cross-hand phase (RL/LR) displayed a non-linear response, and residual phase ramps persisted regardless of the choice of calibrator sources. An engineering experiment by \citet{2025rai} revealed that the uGMRT signal chain introduces a variable cross-phase response depending on the fractional polarization. This behaviour undermines the usual assumption that calibration can be transferred between sources, leading to significant inaccuracies in the polarization calibration. 

These instabilities are particularly problematic because they mimic astrophysical effects such as Faraday rotation and depolarization, making it difficult to disentangle true source polarization from instrumental artifacts. In this paper, we present a systematic investigation of this polarization calibration problem and demonstrate that the Quadrature Hybrid (QH) polariser in the uGMRT Band 4 signal chain is the primary source of the instability. We describe a controlled experiment in which the QH polariser was bypassed in a subset of antennas, converting them from circular to linear polarization feeds, and show that this modification results in stable cross-hand phases, dramatically reduced leakage amplitudes, and accurate recovery of polarization angles and rotation measures.

The paper is structured as follows. Section~\ref{sec:sigchain} presents the architecture of the uGMRT signal chain. Section~\ref{crisis} discusses the polarization calibration challenges encountered with the standard system and details the engineering investigations that identified the QH polariser as the source of the problem. Section~\ref{sec:experiment} describes a controlled astronomical experiment conducted with bypassed polarisers. Section~\ref{sec:calibration} outlines the data analysis methodology applied to both circular and linear feed configurations. Section~\ref{sec:results} reports the comparative results obtained from the two systems. Section~\ref{sec:another} presents a recent, simplified noise-source experiment comparing the same systems. Finally, Section~\ref{sec:conclusions} summarizes our conclusions and provides recommendations for achieving reliable sub-GHz polarimetry with the uGMRT.

\begin{figure*}[ht!]
\includegraphics[width=\textwidth]{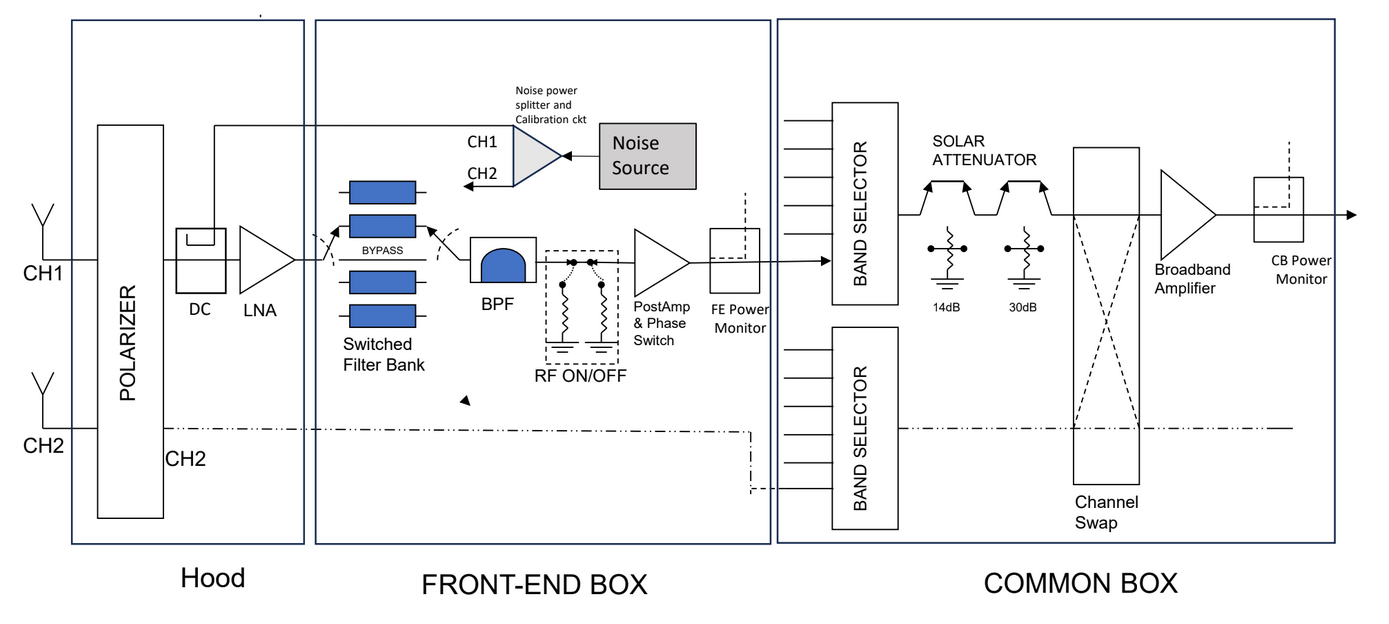}
\caption{Block diagram of the uGMRT dual-channel RF signal chain for a single antenna. CH1 and CH2 are the polarization inputs from the feed. The system is divided into three sections: Hood (polariser and LNA), Front-End Box (switched filter bank, band-pass filter, RF on/off, post-amplifier with phase switch, and noise source with calibration), and Common Box (band selectors, solar attenuators, channel swap, broadband amplifier, and combined power monitor). The Front-End and Common Box are located at the antenna turret just behind the feeds}
\label{frontend}
\end{figure*} 

\section{\MakeLowercase{u}GMRT Signal Chain}\label{sec:sigchain}
The uGMRT signal chain begins at the prime focus of each 45-meter antenna, where wideband feeds capture radiation across four frequency bands: Band 2 (120--250 MHz), Band 3 (250--500 MHz), Band 4 (550--950 MHz), and Band 5 (1000--1460 MHz) \citep{2017CSci..113..707G}. For Bands 2--4, the linear polarization signals from the feeds are converted to circular polarization using a QH (Figure \ref{frontend} ) polariser before reaching the low-noise amplifiers (LNAs), while Band 5 maintains linear polarization throughout the signal chain. Each polarization channel has a dedicated LNA with noise injection capabilities (Figure \ref{frontend}, at four selectable power levels for system temperature measurements. The amplified signals from both polarizations pass to a common box mounted on the feed turret, where the desired frequency band is selected for transmission. The common box also provides solar attenuator options (0, 14, 30, or 44 dB), Walsh modulation control, and the ability to swap polarization channels. From the common box, the RF signals travel directly to the antenna base.

In the upgraded system, high dynamic range optical fibres transmit the unmodified RF signals from the antenna base to the Central Electronics Building (CEB), with fibre lengths ranging from approximately 600 meters for the central square antennas to 21 kilometres for the outermost arm antennas \citep{2014ASInC..13..453S}. At the CEB, the GMRT Analogue Backend (GAB) receives and processes the RF signals. The GAB down-converts the selected RF band to baseband, with bandwidth options of 100, 200, or 400 MHz. This frequency conversion employs independently configurable local oscillators for each antenna and polarization, with frequency settings available across 100--1700 MHz in 10 kHz steps. The GAB also includes variable attenuators for power equalization across the array.

The baseband signals are digitized and fed to the GMRT Wideband Backend (GWB, \citet{2017JAI.....641011R}), which implements an FX correlator architecture for interferometric processing. The GWB digitizes the full 400 MHz bandwidth from both polarizations for all 30 antennas, channelizes the data into up to 16384 spectral channels, and computes the full polarization correlation products for all antenna pairs. The GWB can simultaneously form up to four independent beams (incoherent array, phased array, or coherently de-dispersed modes) with independently selectable antenna subsets for each beam. The final correlation products or beam-formed data are time-averaged according to the observing requirements and recorded for subsequent analysis.

\section{The \MakeLowercase{u}GMRT Band 4 Circular Feed and Polarization Experiments}\label{crisis}

The uGMRT Band 4 system operates with natively linear feeds with a polariser which converts them to circularly polarized signals. The choice of converting the native linear feeds to circular feeds has both historical origins and apparent practical advantages in some scenarios. In circular feeds, the instrumental leakages appear in the first or higher order, while in linear feeds the cross-polarization corruptions appear in the zeroth order. This makes the calibration of linear feeds an inherently iterative process. The differential Faraday rotation of the ionosphere is imprinted as a change in phase for circular feeds, while it manifests as amplitude decorrelation in linear feeds. Traditionally, in beam-formed applications such as pulsar observations, only phases were applied, but with linear feeds one must also compensate for amplitudes to achieve coherent addition of antenna signals. Another aspect is calibration errors, which are quite common in radio astronomy. These largely manifest in the polarization state of the feed. Owing to the nature of the feed, circular feeds are more prone to producing errors in the measurement of circular polarization, and similarly linear feeds can introduce errors in the measurement of linear polarization. However, in most interferometric studies, measurements of linear polarization are generally more in demand than those of circular polarization. Historically, considering these trade-offs, telescopes were built with either native circular feeds or linear feeds with a converter. At uGMRT frequencies, the size of a native circular feed and its inherent inability to support a broadband response required the use of a linear feed with a converter to produce circularly polarized signals. In these polarisers, typically one channel is delayed by $\lambda/4$ and then the two channels are added or subtracted to produce the circular signals. This approach works well for narrow bandwidths, but polarisers cannot perform this operation coherently across a large bandwidth as the delays are tuned to a specific wavelength. Historically, radio telescopes operated with small bandwidths, but the demand of large instantaneous bandwidths in current and upcoming wideband interferometers called this design choice into question. 

\begin{figure*}[ht!]
\centering
\begin{subfigure}{0.48\textwidth}
    \centering
    \includegraphics[width=\linewidth]{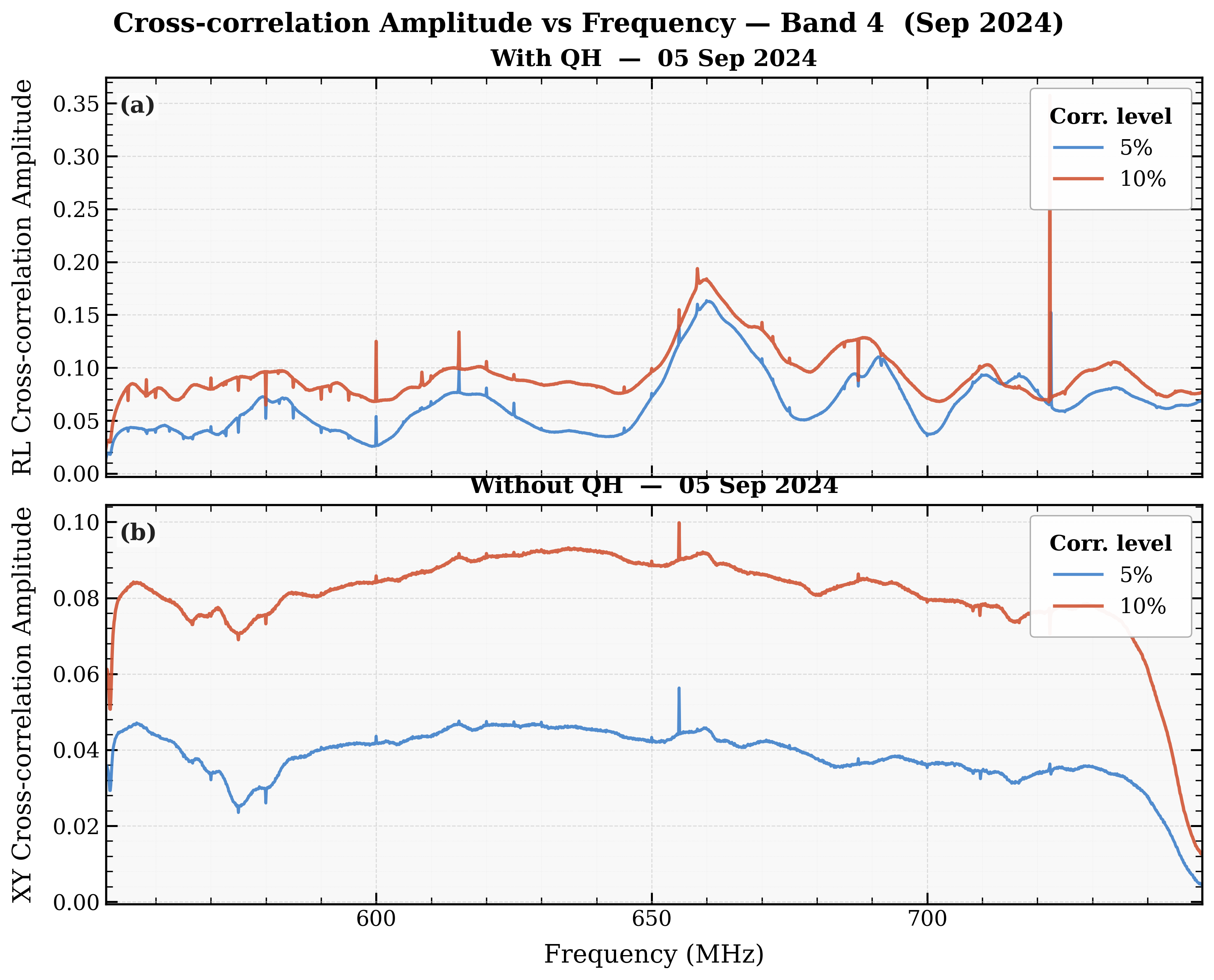}
\end{subfigure}
\hfill
\begin{subfigure}{0.48\textwidth}
    \centering
    \includegraphics[width=\linewidth]{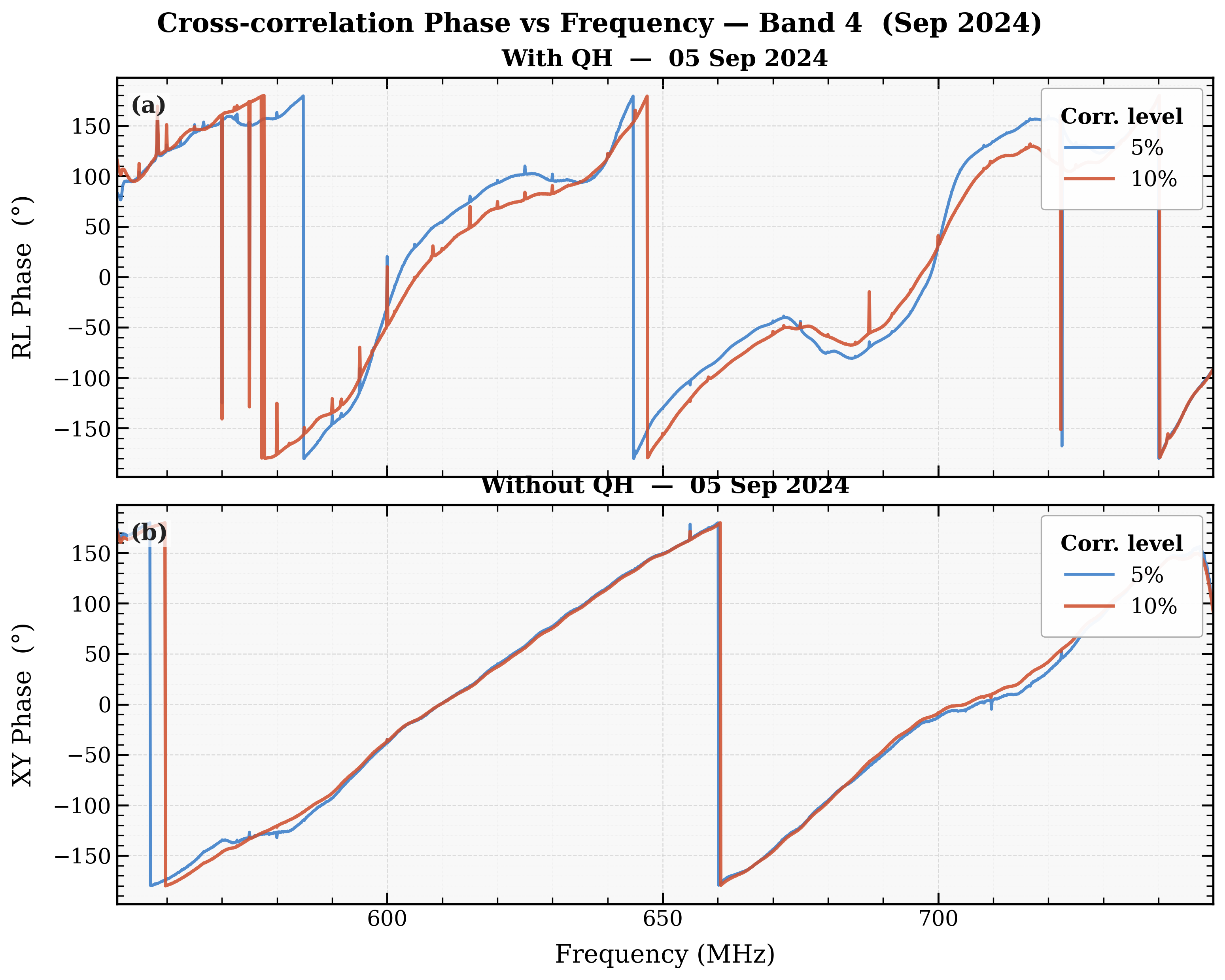}
\end{subfigure}
\caption{Cross-correlation characteristics for two correlation levels (5\% and 10\%) over the 550 to 750 MHz. The left panel illustrates the cross-correlation amplitude as a function of frequency, indicating variations in signal magnitude for the two correlations. The right presents the corresponding phase response, showing the phase evolution. In every case, the top is the response of the system with the QH in place and the bottom is when the QH is bypassed. The figure is adapted from \citep{2025rai}.}
\label{sanjeet_noise}
\end{figure*}

During regular science observation sessions, we identified persistent phase ramps in the cross-hand (RL) visibilities that remained uncorrected even after applying the full polarization calibration chain, including leakage correction, cross-hand delay estimation, and polarization angle calibration. Such residual phase slopes could arise from uncorrected ionospheric Faraday rotation, instrumental delays not captured by the calibration process, or inaccuracies in the calibration solutions themselves. A similar systematic behaviour was also observed in beam-formed data (credits: Dipanjan Mitra), indicating that the effect was not specific to the interferometric correlator mode. Furthermore, the residual phase ramps persisted even when different calibrators were used, suggesting that the origin of the issue is likely intrinsic to the system or propagation effects rather than related to a particular calibration source.

The observed cross-hand phase behaviour necessitated a systematic investigation of the signal chain. At sub-GHz frequencies, the sky contains relatively few bright polarized sources suitable for calibration, which makes diagnosing instrumental effects using astronomical observations alone challenging. Therefore, we chose to first test the signal chain using engineering equipment before proceeding to verification with astronomical observations.

\citet{2025ganla} designed a two-channel correlated noise source with three inputs and two outputs. By adjusting the relative weights of the three input noise sources, this device produces a correlated noise signal that mimics a polarized astronomical signal with controllable fractional polarization. The varying correlation level serves as a proxy for changing the degree of polarization, allowing systematic testing of the instrumental response across a range of polarization states.

The uGMRT signal chain can be divided into five principal blocks: the feed, the frontend system (including the QH polariser and LNAs), the optical fibre link (OFC), the analogue backend (GAB), and the digital correlator (GWB). \citet{2025rai} injected the correlated noise source at the input of each successive block and compared the measured correlations at the correlator output against the expected values. The results demonstrated that the signal path from the OFC onwards is well-behaved, producing negligible leakage and stable cross-phase response with the change of the fractional correlation of the noise source. However, when the noise source was injected at the input of the frontend system, the output exhibited substantial corruption with a variable phase slope that depended on the correlation level of the input signal. Since the cable lengths remained fixed and only the input correlation was varied, the observed slope variation could not be attributed to differential delays. This experiment isolated the problem to the frontend system.

With the frontend system identified as the source of the instability, \citet{2025rai} proceeded to test individual components within it. The frontend system architecture is described in detail in \citet{2025rai}. \citet{2025rai} conducted an engineering test in which a correlated noise source \citep{2025ganla} was connected directly to the frontend inputs, bypassing the feed, while keeping the remainder of the system unchanged. The experiment was performed in two configurations: first with the quadrature hybrid (QH) in place, and then with the QH bypassed. The results (Figure \ref{sanjeet_noise}) showed that, in the presence of the QH, the system exhibited non-linear delays that varied with the strength of polarization or correlation of the injected noise signal. In contrast, when the QH was bypassed, the system behaved as expected, showing a stable response that was independent of the input polarization state. Also, in the amplitude, the with QH system showed oscillating structures while the without QH system correctly showed a flatter profile, showing more stable response. These findings indicated that the observed issue originated from the quadrature hybrid polariser.

\section{A Controlled Experiment}\label{sec:experiment}

The engineering tests identified the QH polariser as the source of the instability, but astronomical observations were required to verify this finding under realistic observing conditions. We designed a controlled experiment in which the QH polariser was bypassed in a subset of antennas, converting them from circular to linear polarization feeds while keeping all other components unchanged.

\begin{figure*}[ht!]
\includegraphics[width=\textwidth]{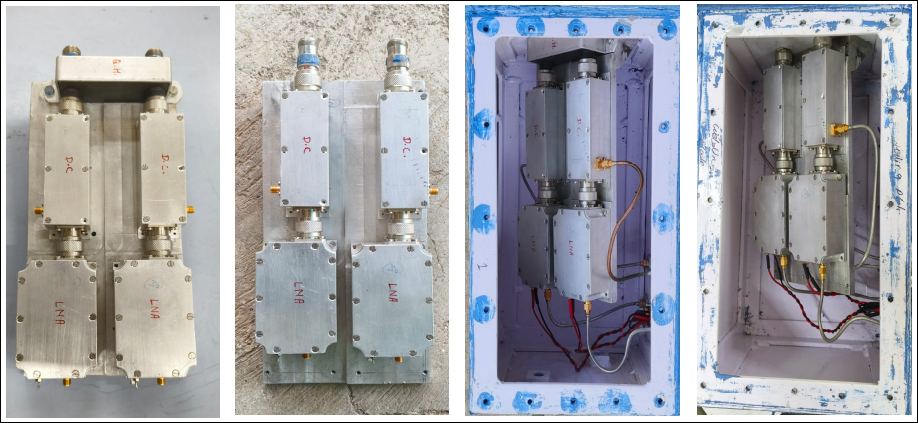}
\caption{Zoomed in version of the uGMRT band 4 feed electronics. The first panel shows, from top to bottom, the QH polariser, the DC coupler and LNA for 2 of the polarization channels. The second panels shows the exact same thing but the QH being removed and again a direct coupler is used to connect to the linear feeds. The third panel shows the regular electronics with QH, coupler and the LNA with all the connections inside the hood. The last panel shows the same thing but the QH being removed.}
\label{hood}
\end{figure*}

Based on the engineering analysis, the polarization corruptions in the uGMRT Band 4 system originate primarily from the feed and the frontend electronics, with the QH polariser responsible for the dominant contribution to the cross-coupling in the frontend system. With the QH removed, only the feed can produce dominant cross-coupling. To establish a baseline for comparison, we first observed a set of calibrator sources over two days with the standard system configuration. These observations provided a template of the instrumental response under normal operating conditions. Following this, the QH polarisers were bypassed in seven antennas using direct couplers, with no other modifications to the signal chain (Figure \ref{hood}). We then repeated the observations on the same sources with identical observing settings.

We typically observed standard unpolarized and polarized calibrator sources like 3C147, 3C286, DA240 and a variety of pulsars with different RM and fractional polarization, for the interferometric analysis, we are going to present only the analysis of interferometric calibrators such as 3C147 (unpolarized), 3C286 (2-5\% fractional polarization strength) and DA240 (25-30\% fractional polarization strength).

The seven antennas selected for the bypass experiment were C04, C08, C09, and C12 from the central square, and W01, E02, and S01, one from each arm of the Y-shaped array. This selection was driven by two considerations. First, the short baselines among these antennas provide adequate sensitivity for point source observations while maintaining a reasonably Gaussian synthesized beam, which is sufficient for our diagnostic purposes since the calibrators are unresolved. Second, the central square antennas are logistically accessible, which was an important practical constraint. The uGMRT feeds are weatherproofed and sealed, so the bypass procedure required removing each feed assembly, transporting it to the laboratory for modification, and reinstalling it on the antenna. Each antenna required multiple operations with the cherry-picker vehicle used for feed access. Concentrating the modifications on nearby antennas minimized the logistical burden on the engineering team.

After the bypass, observations were carried out in both beam-formed and interferometric modes using the standard uGMRT full-polarization setup. In this paper, we focus exclusively on the interferometric visibility analysis, which allows for detailed examination of antenna-based instrumental effects. In both cases, the data were recorded with 2048 spectral channels across a 200 MHz bandwidth (550–750 MHz) and a time integration of 5.3 s. The same command files were executed for both sets of observations, with only the starting dates adjusted to ensure that the with QH and without QH runs were as closely matched as possible. The with QH data were collected on 31 March 2025, while the without QH data were taken on 10 April 2025.

\section{Data Analysis Methodology}\label{sec:calibration}

All observed sources are unresolved point sources on our baseline scales, which simplifies the calibration procedure. We describe below the calibration methodology for both datasets: the circular polarization (RL) data from the standard system and the linear polarization (XY) data from the QH-bypassed antennas. All data reduction was performed using the Common Astronomy Software Applications package \citep[CASA 6.5;][]{2022PASP..134k4501C}.

\subsection{Calibration of Circular Polarization Data (With QH)}

For the standard circular feed configuration, 3C147 and 3C286 served as the primary calibrators for gain, delay, and bandpass estimation. The calibration proceeded as follows. First, obviously corrupted data were flagged manually, followed by automated flagging using the CASA task \texttt{flagdata} in \texttt{tfcrop} mode with conservative thresholds of $6\sigma$ in both time and frequency to remove outliers.

Initial gain phases were estimated from the full scans of 3C147 and 3C286. Using these initial solutions, parallel-hand residual delays were determined across the full bandwidth using the CASA task \texttt{gaincal} with \texttt{gaintype=`K'}. The bandpass response was then computed from both calibrator scans using the CASA task \texttt{bandpass}. With the bandpass applied, refined gain amplitudes and phases were estimated using \texttt{gaincal} with \texttt{gaintype=`G'} and \texttt{calmode=`ap'}.

For polarization calibration, we note that the uGMRT Band 4 system produces a flipped handedness due to reflection by the parabolic reflector \citep{2020arXiv200408542D}. To correct for this, we implemented a header swap, manually reconfiguring the Stokes keywords from RR, RL, LR, LL to LL, LR, RL, RR before proceeding with polarization calibration.

Cross-hand delays were estimated using the CASA task \texttt{gaincal} with \texttt{gaintype=`kcross'} on the 3C286 scan. Instrumental polarization leakage was determined from the 3C147 scan, assuming it to be unpolarized in the 550--750~MHz range, using the CASA task \texttt{polcal} with \texttt{poltype=`Df'}. Finally, the cross-hand phase corrections were computed on 3C286 using \texttt{polcal} with \texttt{poltype=`Xf'}, assuming a full polarization model for 3C286. We note that extrapolation of higher frequency ($>1$~GHz) polarization measurements to sub-GHz frequencies does not accurately capture the polarization properties of 3C286, hence an updated model for these frequencies have been supplied for calibration (personal communications with Preshanth Jagannathan, Ben Hugo \citep{hugo2024absolute} and \citet{2026ApJS..283...82P}).

\begin{figure*}[htbp!]
    \centering
    \begin{tabular}{ccc}
        \includegraphics[width=0.3\textwidth]{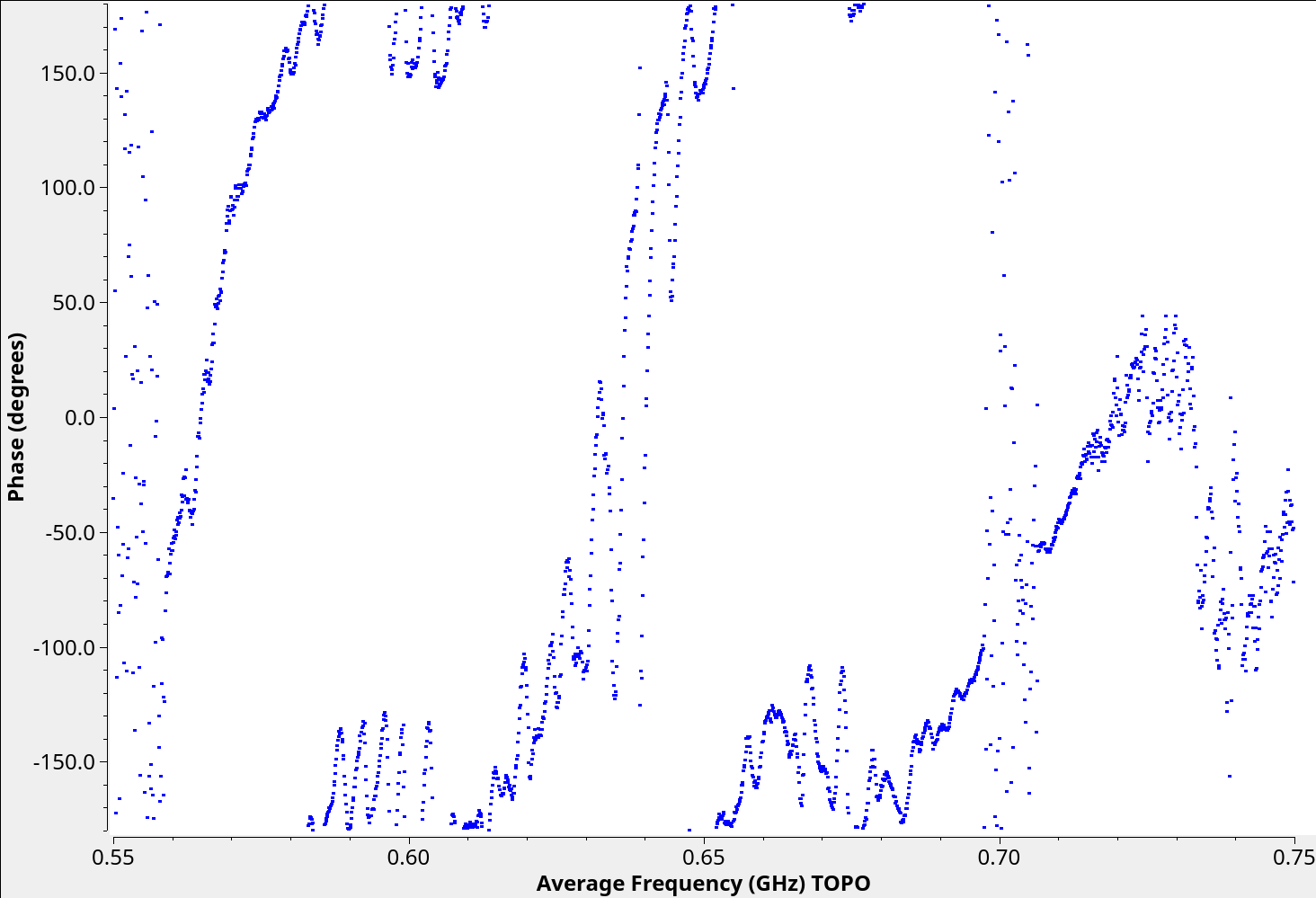} &
        \includegraphics[width=0.3\textwidth]{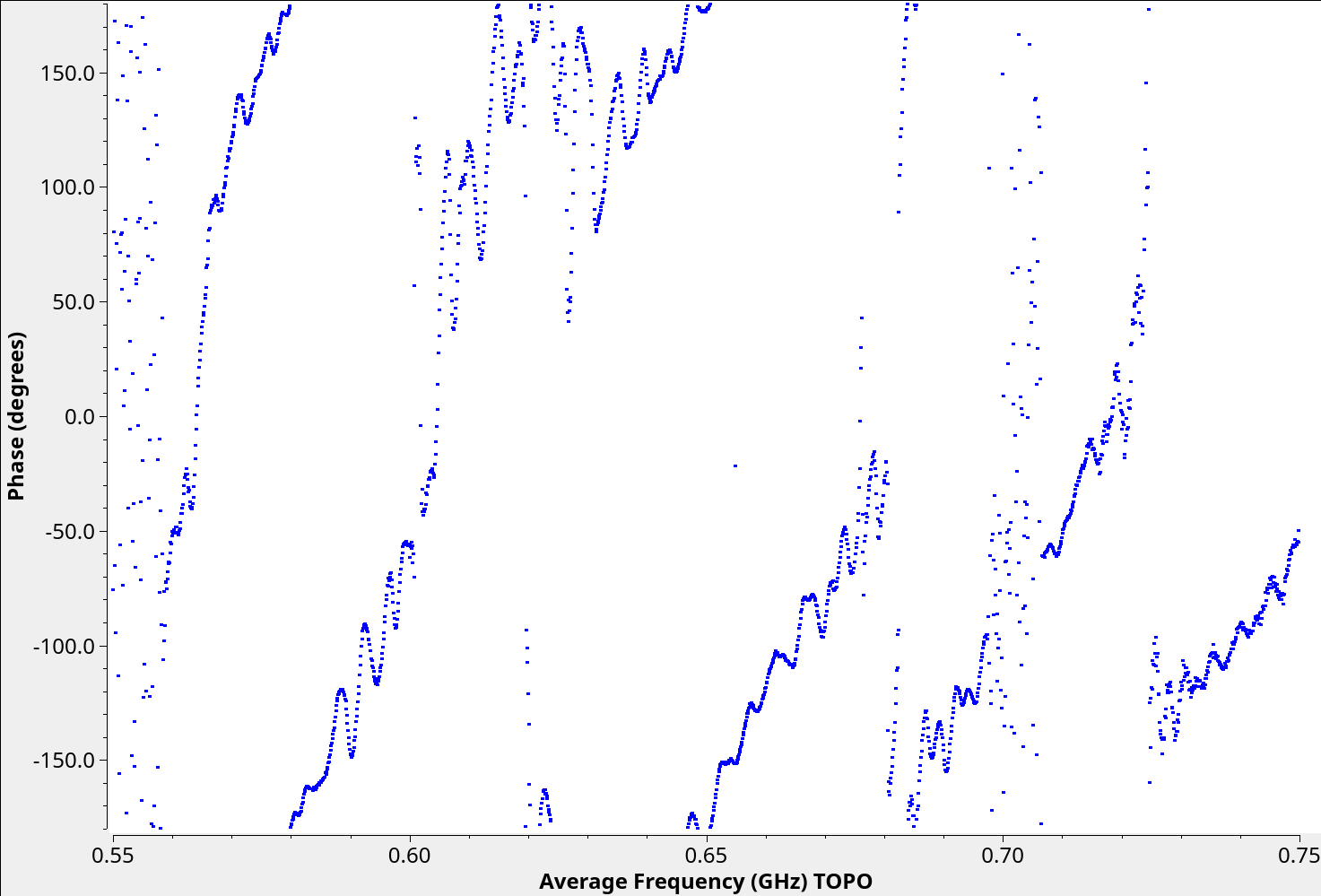} &
        \includegraphics[width=0.3\textwidth]{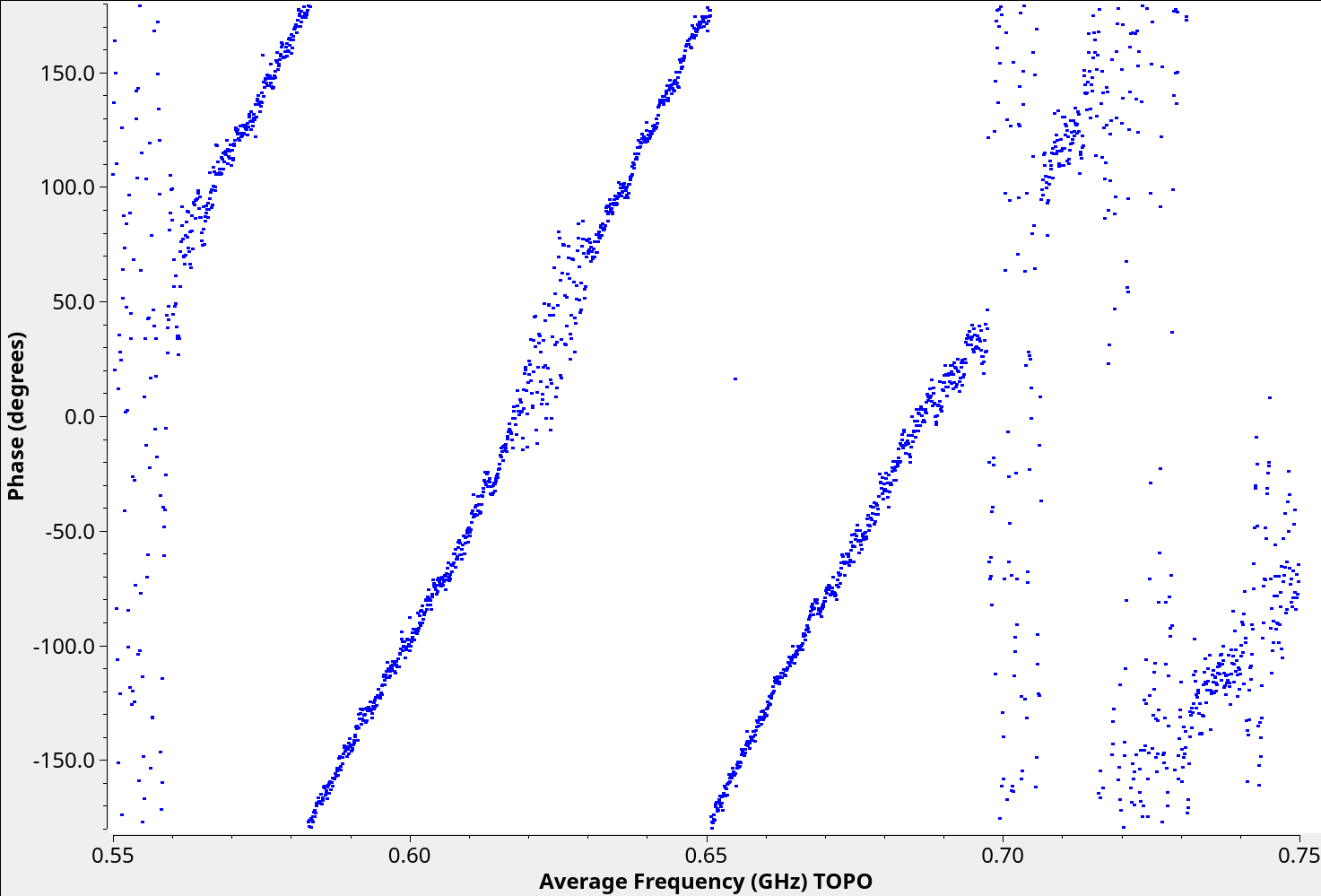} \\
        3C147 & 3C286 & DA240 \\[0.5cm]
        \includegraphics[width=0.3\textwidth]{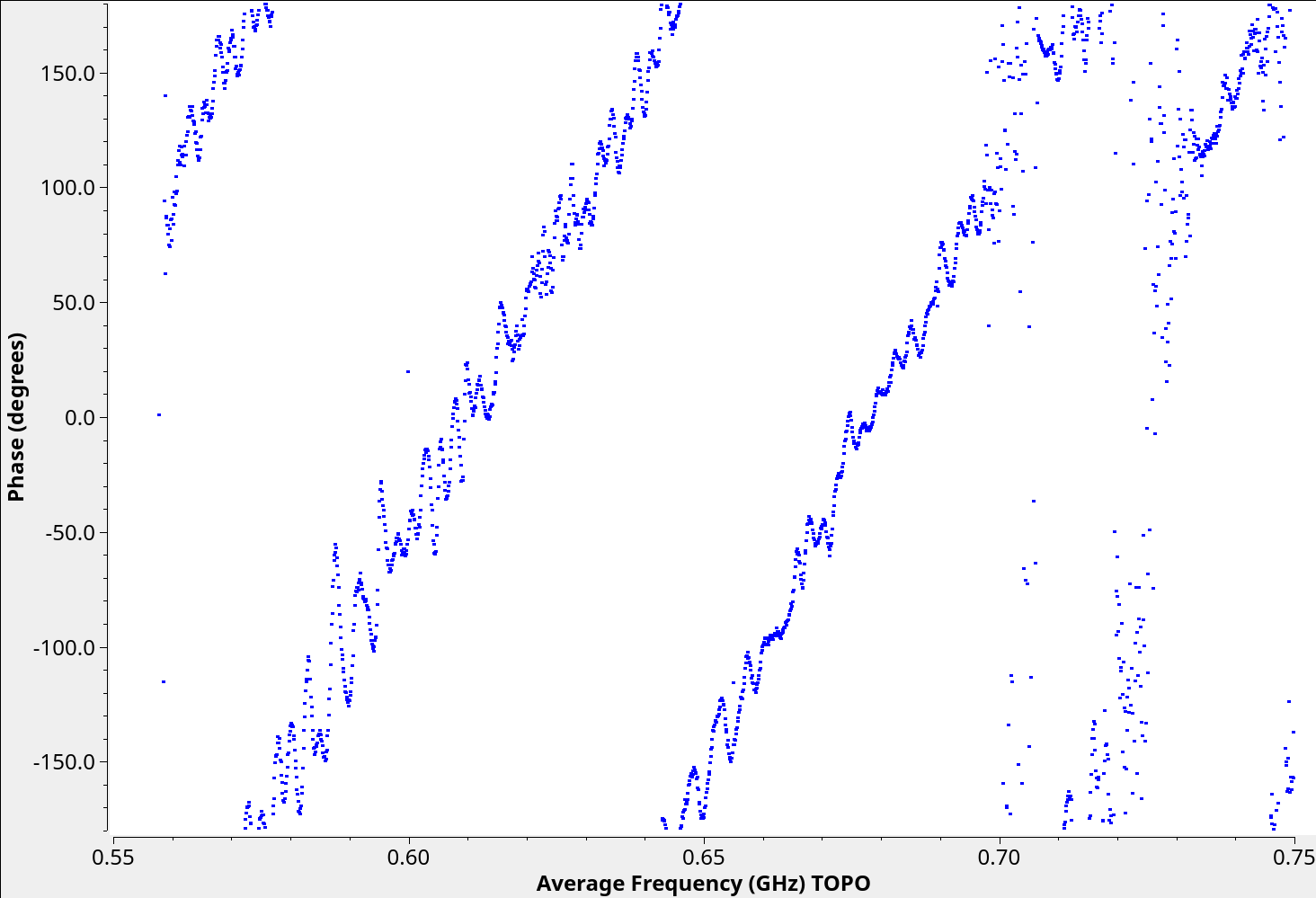} &
        \includegraphics[width=0.3\textwidth]{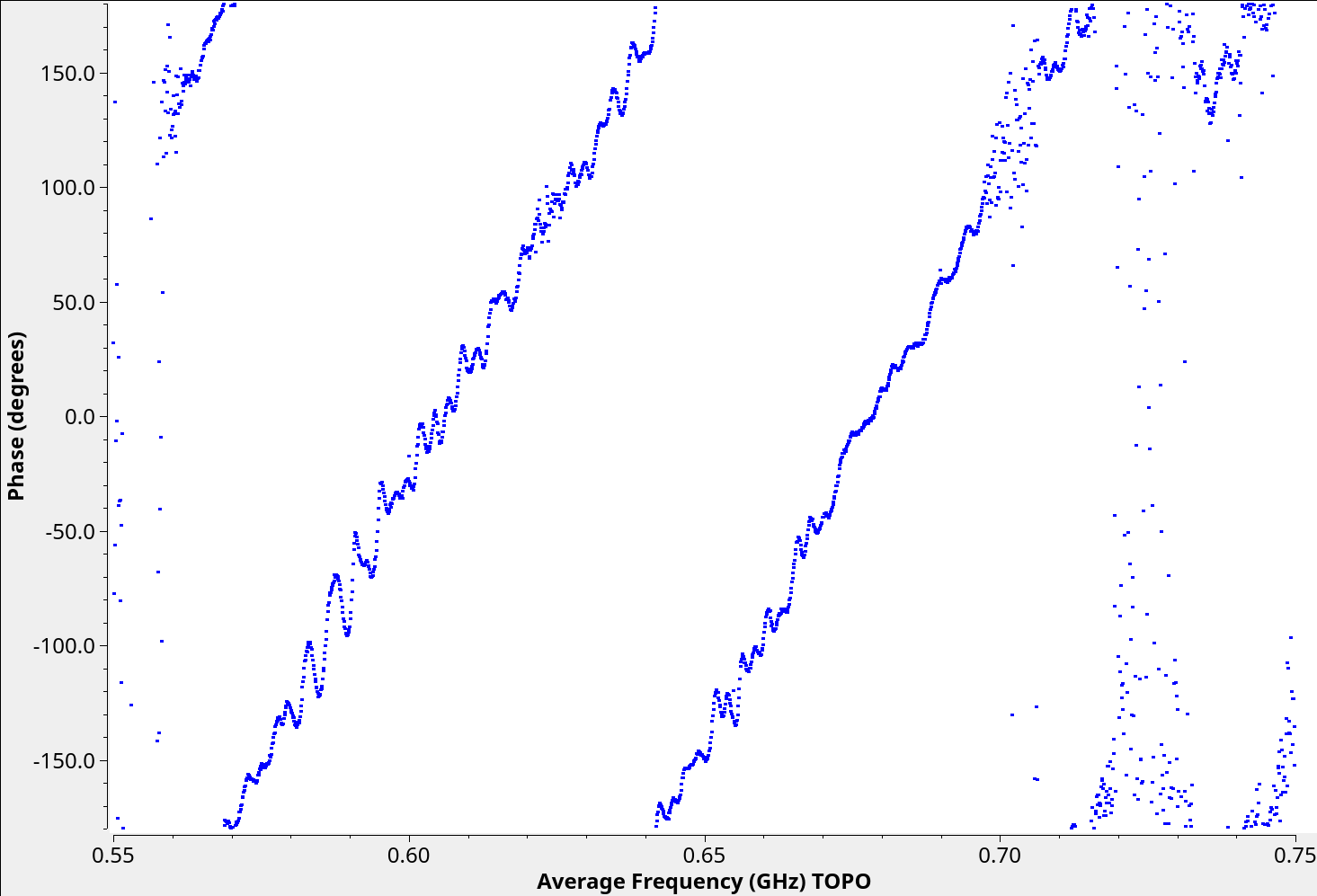} &
        \includegraphics[width=0.3\textwidth]{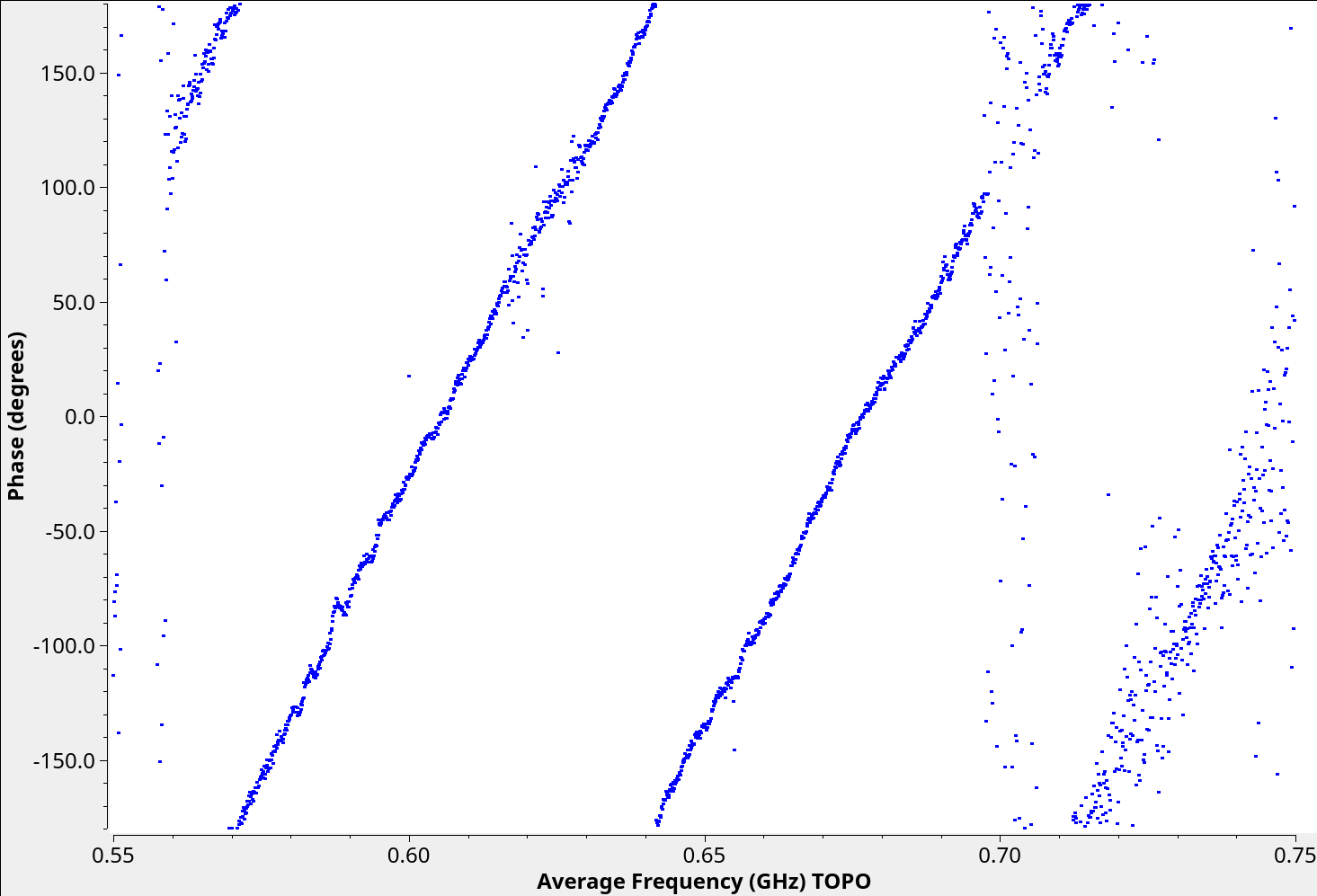} \\
        3C147 & 3C286 & DA240\\
    \end{tabular}
    \caption{The variation of the raw cross-hand (RL and XY) phase with frequency. The upper panel shows the phases for 3 sources with different polarization properties, unpolarized (3C147), weakly polarized (3C286) and strongly polarized (DA240). The exact same is shown in lower panel but with QH bypass in the GMRT signal chain. For all of the panels, the y-axis varies from -180 to 180 degrees and the x-axis has limits from 550 - 750 MHz.}
    \label{qh_nqh_raw_phase}
\end{figure*}

\subsection{Calibration of Linear Polarization Data (QH Bypassed)}

For the QH-bypassed data recorded in linear (XY) polarization basis, the calibration procedure differs in several important respects. In the XY basis, the parallactic angle rotation couples the gains and leakage terms, requiring an iterative approach to obtain accurate solutions.

Initial bandpass and gain estimates were derived from the 3C147 scan. With these applied, the instrumental leakage terms were estimated. A full polarization model for 3C286 was then imported, and gains for 3C286 along with cross-hand phase corruptions were computed. To resolve the $\pi$ ambiguity inherent in cross-hand phase calibration for linear feeds, the CASA task \texttt{xyamb} was employed. For proper gain calibration in the XY basis, we used \texttt{gaintype=`T'} to preserve the relative power levels between the XX and YY correlations, which is essential for accurate Stokes parameter reconstruction in linear polarization systems.

All calibration solutions were applied to the complete dataset for subsequent analysis. In both the cases, C04 was used as a reference antenna.

\section{Results of the Experiment}\label{sec:results}

We compare the raw and calibrated data from the with QH and QH-bypassed configurations.

\subsection{Raw Data Comparison}

Figure~\ref{qh_nqh_raw_phase} shows the raw cross-hand phases (RL for circular feeds, XY for linear feeds) for the average of the baselines to C04 as a function of frequency for three sources with different polarization properties at our observing frequencies: 3C147 (unpolarized), 3C286 (weakly polarized, $\sim$2-5\%), and DA240 (strongly polarized, $\sim$20-30\%). The upper panels show the standard system with QH, and the lower panels show the QH-bypassed configuration.

The difference between the two configurations is striking. In the with QH system, the cross-hand phase varies dramatically with the polarization strength of the observed source. The phase structure is non-linear across frequency, inconsistent with a simple differential delay. This behaviour has severe implications for polarization calibration. Leakage calibration typically employs an unpolarized source, but the phase response of the QH depends on the polarization state of the input signal. Although a polarized calibrator is subsequently used to determine the absolute cross-hand phase, the source-dependent response means that corrections derived from one calibrator cannot be reliably transferred to a target source with different fractional polarization.

The middle column shows 3C286, the most widely used polarization calibrator at low frequencies. With the QH in place, the cross-hand phase spectrum is non-linear and cannot be adequately described by a simple delay. This also validates the discrepancy observed in our diagnostic experiment (Section \ref{crisis}). 3C286 and DA240 have different fractional polarizations and therefore produce different instrumental responses. When calibration solutions from one source are applied to the other, the residual phase errors corrupt the target data. This limitation implies that with the standard with QH system, accurate polarization calibration is only achievable when the calibrator and target have comparable fractional polarization. It is worth noting that for a higher fractional polarization, the cross-phase shows a linear behaviour (Figure \ref{qh_nqh_raw_phase}, DA240 panel). 

In contrast, the QH-bypassed system exhibits completely stable cross-hand phases. The observations span 7 hours, with 3C147 observed at the beginning and DA240 at the end of the session. Throughout this period, the cross-hand phase remains constant and independent of the source polarization. This behaviour conforms to the standard assumptions of radio interferometric calibration, that corrections derived on a calibrator can be transferred to any target source regardless of its polarization properties.
\begin{figure*}[ht!]
    \centering
    \begin{subfigure}{0.49\textwidth}
        \includegraphics[width=\textwidth]{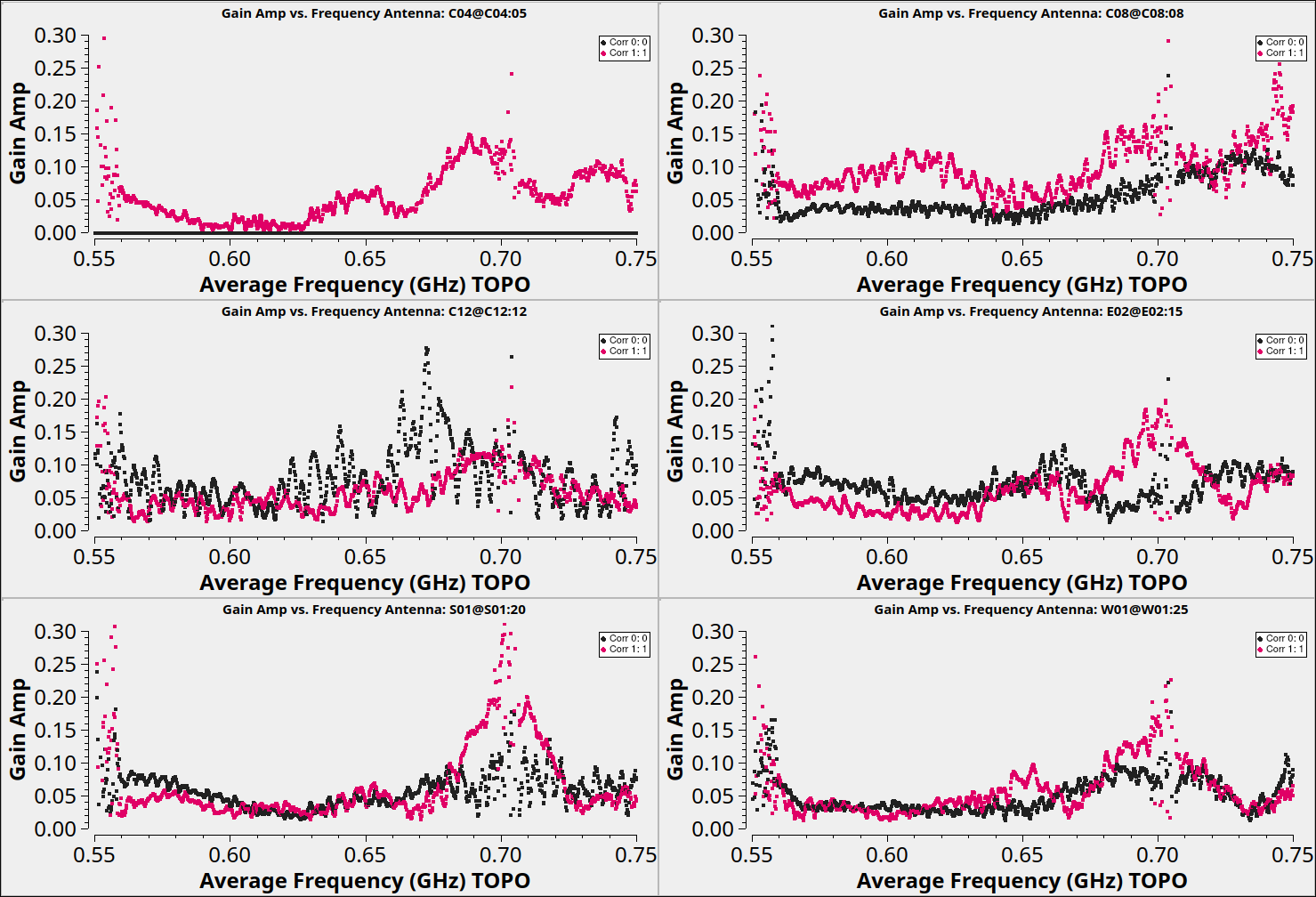}
    \end{subfigure}
    \hfill
    \begin{subfigure}{0.49\textwidth}
        \includegraphics[width=\textwidth]{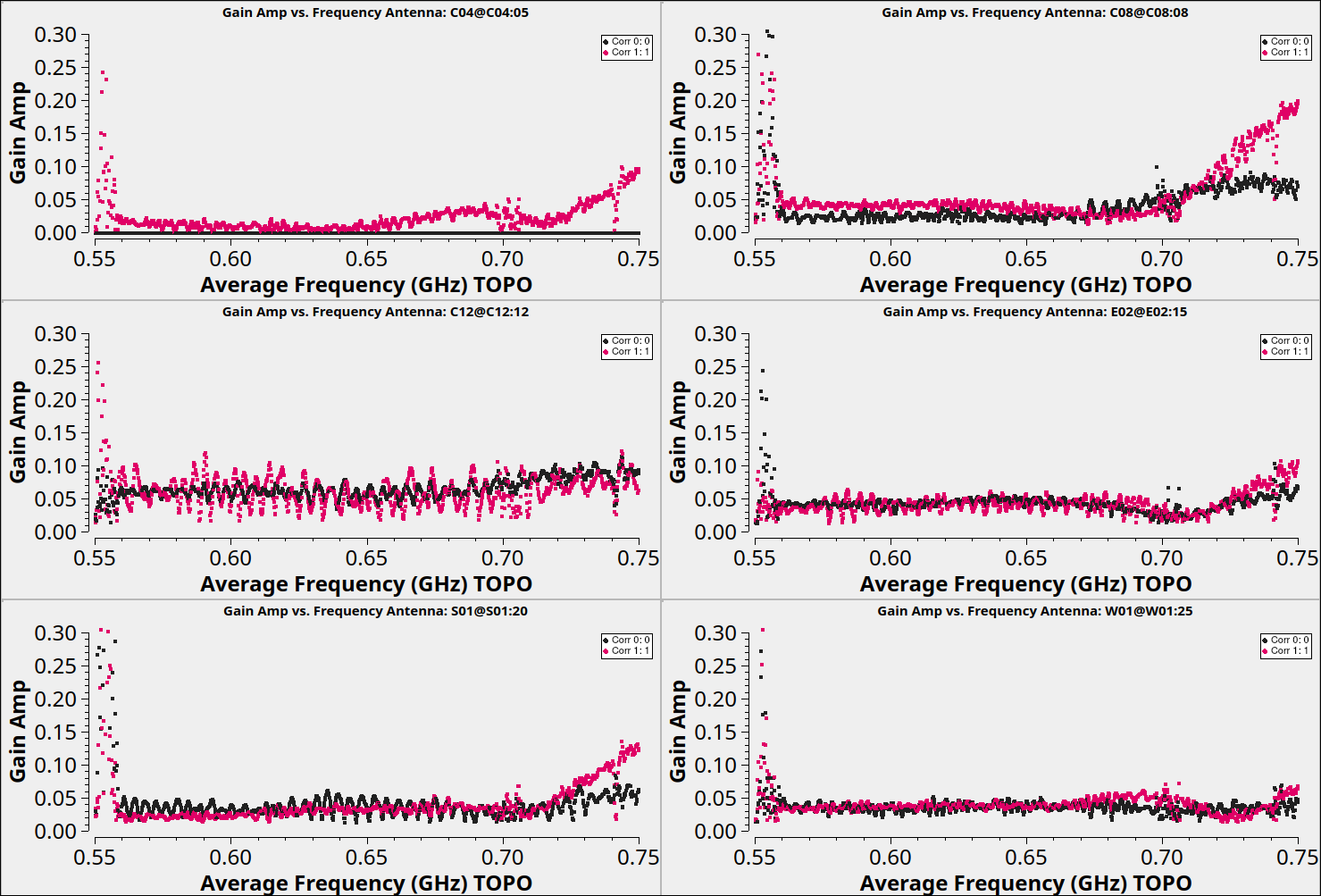}
    \end{subfigure}
    \caption{Leakage amplitude comparisons are shown for the cases with the QH and with the QH bypassed. The Y-axis represents the relative leakage in fractional units, and the X-axis shows frequencies from 550 to 750 MHz. The same set of antennas, C04, C08, C12, E02, S01, and W01, is shown in separate panels. The black and red colours indicate the polarization products: RL/LR and XY/YX, corresponding to the respective panels.}
    \label{leak_compare}
\end{figure*}

\subsection{Difference in Cross-coupling}

Figure~\ref{leak_compare} compares the leakage amplitudes for the with QH and QH bypassed configurations. The difference is substantial. In the standard system, the QH polariser introduces significant cross-coupling between the polarization channels, resulting in average leakage amplitudes of 10--15\%. In the bypassed configuration, where the only source of polarization mixing is the feed itself, the leakages are reduced to 2--5\%. A leakage amplitude that is flat across frequency is indicative of a geometric misorientation between the two dipole feeds rather than a frequency-dependent instrumental effect. This is because a physical misalignment of the dipoles is an achromatic property — the angle between the feeds does not change with frequency — and therefore its contribution to the cross-hand signal appears as a constant offset across the band.

To see this formally, consider two nominally orthogonal dipoles where one is 
rotated by a small angle $\varepsilon$ from its ideal orientation. The resulting 
Jones matrix takes the form of a rotation matrix:

\begin{equation}
    J = \begin{pmatrix} \cos\varepsilon & \sin\varepsilon \\ 
    -\sin\varepsilon & \cos\varepsilon \end{pmatrix}
\end{equation}

Comparing the off-diagonal terms with the standard D-term representation of the Jones matrix, where the leakage terms appear as $D_X = \sin\varepsilon$ and $D_Y = -\sin\varepsilon$, it follows directly that $|D| = \sin\varepsilon$ and therefore $\varepsilon = \arcsin(|D|)$. Using the observed leakage value of 2\% ($|D| = 0.02$), the implied misorientation angle is $\varepsilon = \arcsin(0.02) \approx 1.15^{\circ}$.

This is a small but non-negligible misalignment. It should be noted that this estimate assumes the D-terms are dominated purely by a geometric rotation of the feeds.

Beyond the amplitude difference, the spectral characteristics of the leakage also differ markedly. With the QH in place, the leakage spectra exhibit strong modulations across frequency, and the response varies considerably from antenna to antenna. This resembles the finding of \citet{2025rai}. In contrast, the QH-bypassed system produces leakage spectra that are largely flat across the bandwidth. An increase in leakage amplitude is observed above 700~MHz in the bypassed case, which can be most likely attributed to the feed response itself. Nevertheless, the fundamental improvement in stability and reduced leakage amplitude is evident. The remaining calibration tables (gains, bandpass, delays) do not show diagnostic differences between the two configurations and are therefore not discussed further.

\subsection{Difference in Calibrated Data}

Following the full polarization calibration procedure described in Section~\ref{sec:calibration}, we compare the calibrated data from the two configurations. The differences are consistent with expectations based on the raw data analysis.

\begin{figure*}[ht!]
    \centering
    \begin{tabular}{ccc}
        \includegraphics[width=\textwidth]{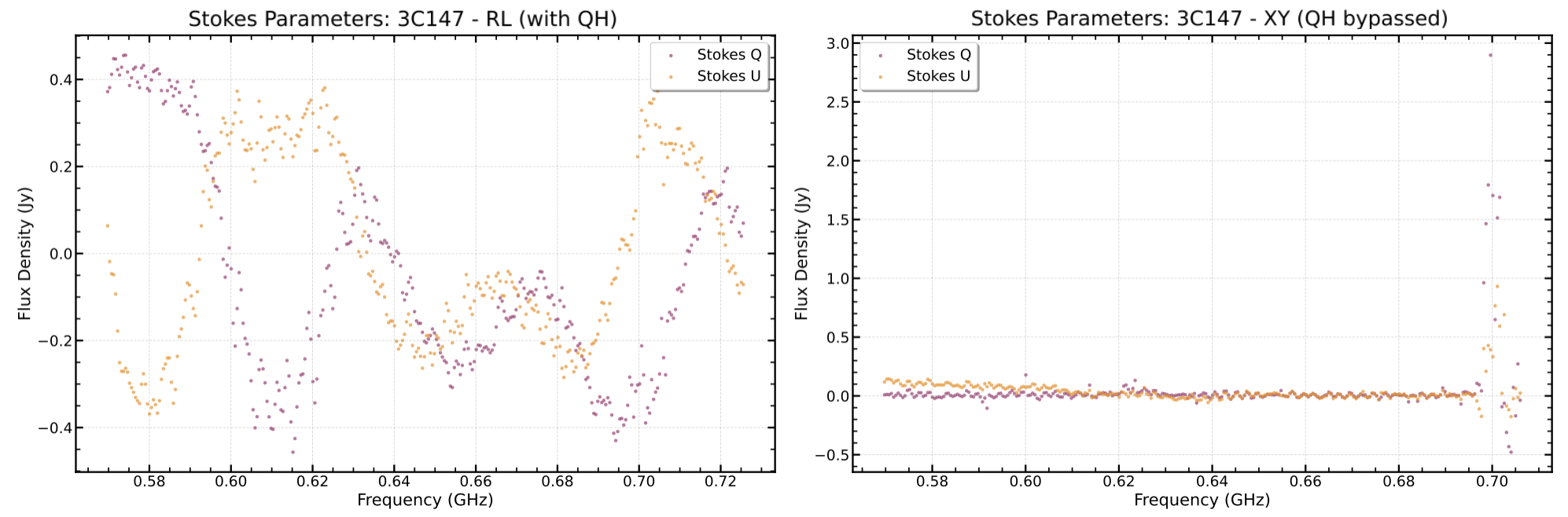}\\
        \includegraphics[width=\textwidth]{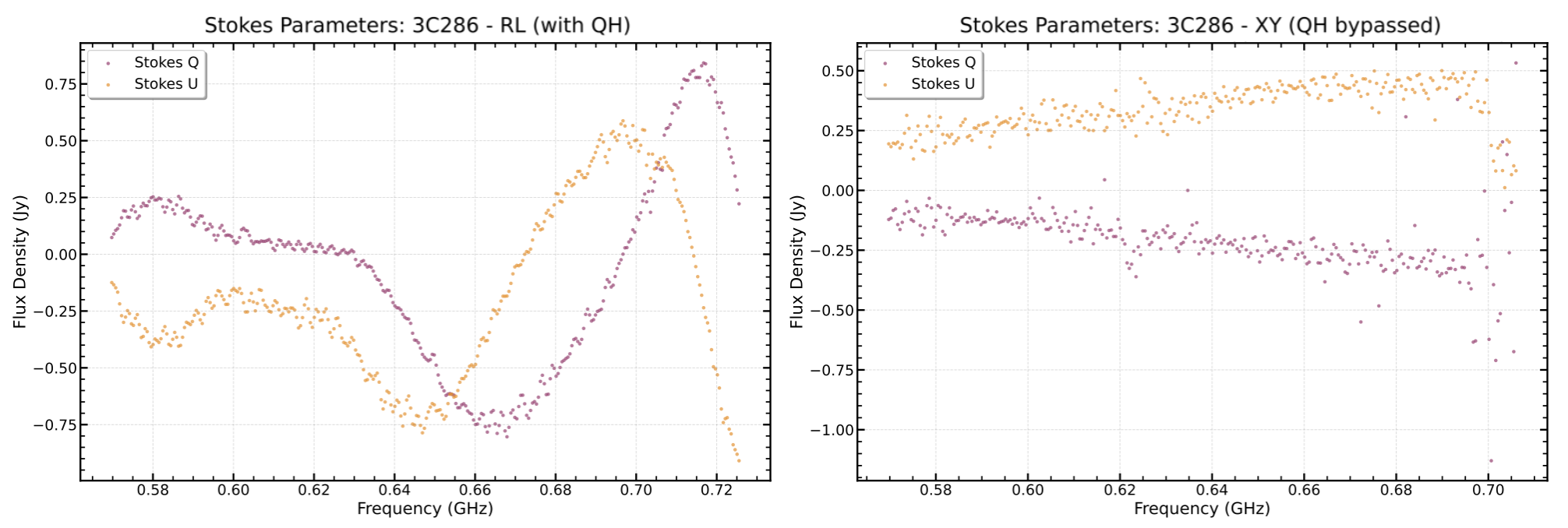} \\
        \includegraphics[width=\textwidth]{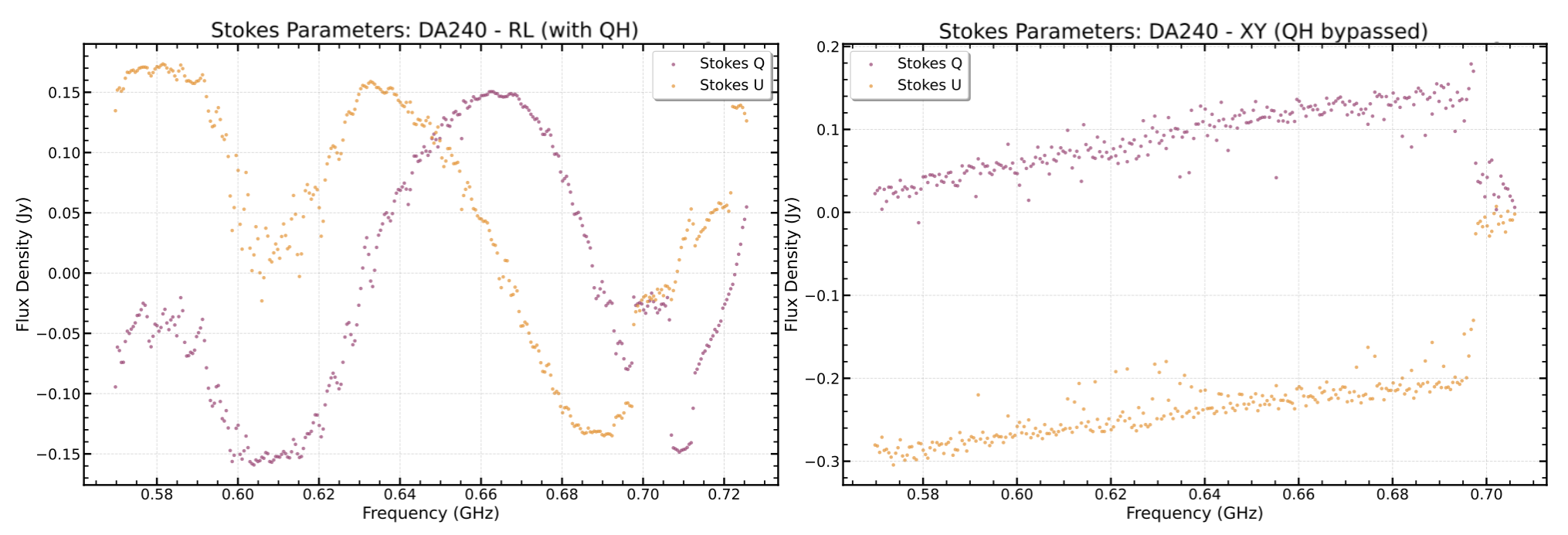} \\
    \end{tabular}
    \caption{The illustration shows the Stokes Q and U spectra for three different sources: 3C147 (top), 3C286 (middle), and DA240 (bottom). In all cases, Stokes Q and U are shown in violet and orange, respectively. For each source, the left panel corresponds to the case with QH, while the right panel corresponds to the case without QH.}
    \label{all_source_compare}
\end{figure*}

We begin with the unpolarized source 3C147, for which perfect calibration should yield zero Stokes Q and U. Figure~\ref{all_source_compare} (top row) shows that in the standard QH system, the residual Stokes Q and U amplitudes are of order $\pm 0.4$~Jy with oscillatory spectral structure. In contrast, the QH-bypassed data show residual amplitudes of approximately 0.1~Jy at the low-frequency end of the band with no oscillatory behaviour. These residuals correspond to leakage levels of $\sim$0.5\% for the standard system and $<$0.2\% for the bypassed configuration, indicating the improved calibration accuracy achievable without the QH polariser.

For the polarization calibrator 3C286 (Figure~\ref{all_source_compare}, middle row), the standard system exhibits residual spectral artefacts even though 3C286 was used to derive the polarization calibration solutions. These artefacts are absent in the QH-bypassed data, where the corrected Stokes Q and U spectra match the expected model. This behaviour is explained by the source-dependent response of the QH. The leakage solutions derived from the unpolarized source 3C147 do not accurately describe the instrumental response when applied to the polarized source 3C286, resulting in residual spectral structure.

\begin{figure*}[htbp!]
    \centering
    \begin{tabular}{ccc}
        \includegraphics[width=\textwidth]{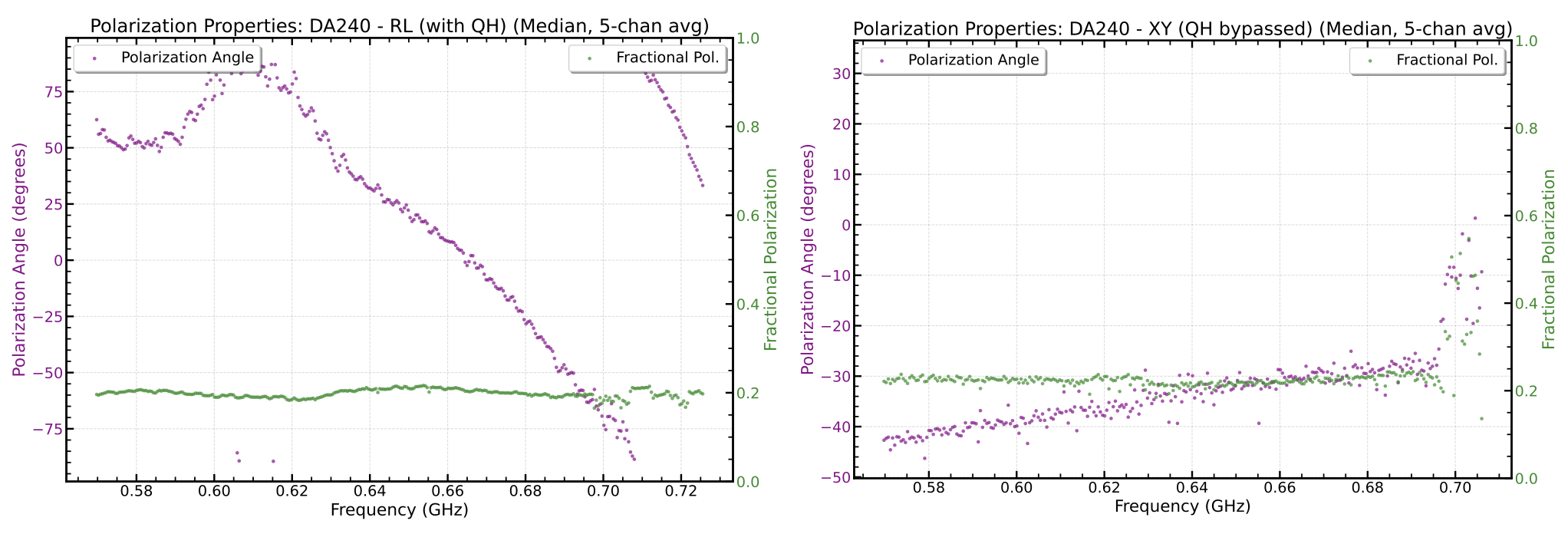}\\
        \includegraphics[width=\textwidth]{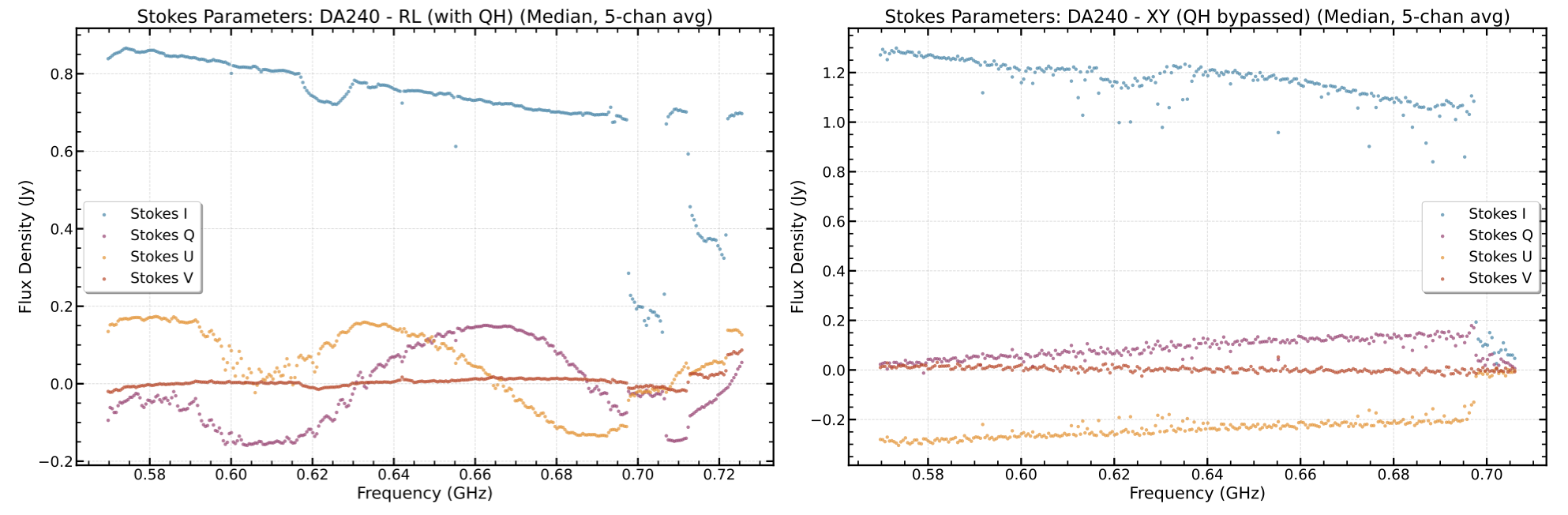} \\
    \end{tabular}
    \caption{The top panels show the fractional polarization and polarization angle spectra for DA240 with QH (left) and without QH (right). In both cases, the fractional polarization is shown in green, while the polarization angle is shown in violet. The bottom panels display the Stokes I, Q, U, and V spectra for DA240, shown in blue, green, magenta, and orange, respectively, with the left panel corresponding to the case with QH and the right panel to the case without QH.}
    \label{DA240_fc}
\end{figure*}

The most significant differences appear in the science target DA240 (Figure~\ref{all_source_compare}, bottom row). The standard system again produces oscillatory residuals in the Stokes Q and U spectra, while the QH-bypassed data are smooth. Figure~\ref{DA240_fc} compares the fractional polarization and polarization angle spectra for both configurations. The fractional polarization is consistent between the two systems and constant across the band. However, the polarization angle behaviour differs dramatically. In the standard system, the polarization angle spectrum is irregular and does not show the expected Faraday rotation signature. In the QH-bypassed system, the polarization angle exhibits a smooth linear slope across the band. DA240 has a rotation measure of 3.3~rad~m$^{-2}$ \citep{2008A&A...489...69B}, which corresponds to a polarization angle rotation of approximately 25$^\circ$ across the 550--750~MHz band. This expected rotation is accurately recovered in the bypassed configuration.

The Stokes I and V spectra are comparable between the two systems, with Stokes V consistent with zero as expected for these sources. The critical difference lies in the Stokes Q and U recovery. These results demonstrate that the standard QH system can measure fractional polarization with residual leakage contamination at the $\sim$0.5\% level, but cannot reliably determine polarization angles or rotation measures. The QH-bypassed system achieves residual leakage below 0.2\% and accurately recovers both the fractional polarization and the Faraday rotation signature.

\subsection{Expected Implications for Stokes I Sensitivity}\label{sec:sensitivity}

Although the primary focus of this work is on polarimetric fidelity, the 
reduced leakage and stable cross-hand phase in the QH-bypassed system are 
also expected to improve total intensity (Stokes~$I$) performance. We 
outline the relevant algebra below, following from the measurement equation 
framework already presented in Section~\ref{sec:sigchain} 
\citep{1996A&AS..117..137H, 1996A&AS..117..149S, 2011A&A...527A.106S}.

The Jones matrix for antenna $i$ can be written as:
\begin{equation}
\mathbf{J}_i = \mathbf{G}_i \mathbf{D}_i = 
\begin{pmatrix} g_{ip} & 0 \\ 0 & g_{iq} \end{pmatrix}
\begin{pmatrix} 1 & d_{ip} \\ d_{iq} & 1 \end{pmatrix},
\end{equation}
where $g_{ip}$, $g_{iq}$ are the complex gains and $d_{ip}$, $d_{iq}$ are 
the leakage terms for the two polarization channels $p$ and $q$.

Consider an unpolarized point source with flux density $S$, for which the 
sky coherency matrix is $\mathbf{V}_{ij}^{\mathrm{sky}} = 
\frac{S}{2}\begin{pmatrix} 1 & 0 \\ 0 & 1 \end{pmatrix}$. From Equation~2, 
expanding the observed parallel-hand visibility $V_{ij}^{pp}$ and 
retaining terms to second order in the leakage:
\begin{equation}\label{eq:vpp}
V_{ij}^{pp} = g_{ip}\, g_{jp}^{*}\, \frac{S}{2} 
\left(1 + d_{ip}\, d_{jp}^{*}\right).
\end{equation}

The Stokes~$I$ visibility is estimated from the sum of the two 
parallel-hand correlations. After perfect gain calibration, the fractional 
error in the Stokes~$I$ estimate due to uncorrected leakage is:
\begin{equation}\label{eq:Ibias}
\frac{\delta I}{I} = \frac{1}{2}\,\mathrm{Re}
\left(d_{ip}\, d_{jp}^{*} + d_{iq}\, d_{jq}^{*}\right).
\end{equation}

This is a second-order term, but it is not negligible for the with-QH 
system where $|d| \sim 0.1$--$0.15$ (Section~\ref{sec:results}), 
corresponding to a per-baseline bias of up to $\sim$1--2\%. Crucially, 
when the leakage spectra exhibit the strong frequency-dependent modulations 
seen in the with-QH configuration (Figure~\ref{leak_compare}), these errors 
do not average out over frequency and introduce spectral artefacts into 
the Stokes~$I$ image.

A separate but equally important effect arises from the cross-hand phase 
instability. Standard self-calibration solves for antenna-based gains 
assuming a diagonal Jones matrix, i.e., $d_{ip} = d_{iq} = 0$. When the 
true off-diagonal terms are large and frequency-dependent, the gain solver 
partially absorbs the leakage into the diagonal solutions. As shown by 
\citet{2001A&A...375..344B}, polarization leakage introduces closure 
errors in the co-polar visibilities that cannot be removed by diagonal 
gain calibration alone. These closure errors manifest as artefacts in the 
Stokes~$I$ image, reducing the achievable dynamic range.

The situation is further compounded when the cross-hand phase varies with 
the polarization state of the observed source (Section~\ref{crisis}). In 
this case, the effective leakage terms are themselves source-dependent, 
meaning that the closure errors change between the calibrator and the 
target. Self-calibration solutions derived on a calibrator then carry 
systematic biases when transferred to the target field, producing 
direction-dependent errors in the final image.

For the QH-bypassed system, with leakage amplitudes of 2--5\% and 
spectrally flat response (Figure~\ref{leak_compare}), the second-order bias 
(Equation~\ref{eq:Ibias}) is reduced to below $\sim$0.1\%. The stable 
cross-hand phase ensures that the leakage terms do not vary with source 
properties, so that self-calibration solutions transfer reliably between 
calibrator and target. Both effects are expected to yield improved 
Stokes~$I$ sensitivity and imaging dynamic range. A detailed quantitative 
analysis using matched observations will be presented in a forthcoming 
paper (Pal et al., in prep.).

\section{More Simplified Noise-source Experiment}\label{sec:another}
Recently, we conducted a simplified test using an updated correlated noise source based on the design of \citet{2025ganla}. The noise source was connected in the receiver room, and a randomly selected frontend box was used for testing. The signal from the noise source was first passed through the frontend box and common box and then connected directly to the GWB backend. The test was repeated both with and without the QH in the frontend system. While these measurements do not fully replicate real observations, as cable delays and OFC systems are absent, they allow the direct assessment of the QH’s effect.

\begin{figure*}[ht!]
\centering
\begin{subfigure}{0.48\textwidth}
    \centering
    \includegraphics[width=\linewidth]{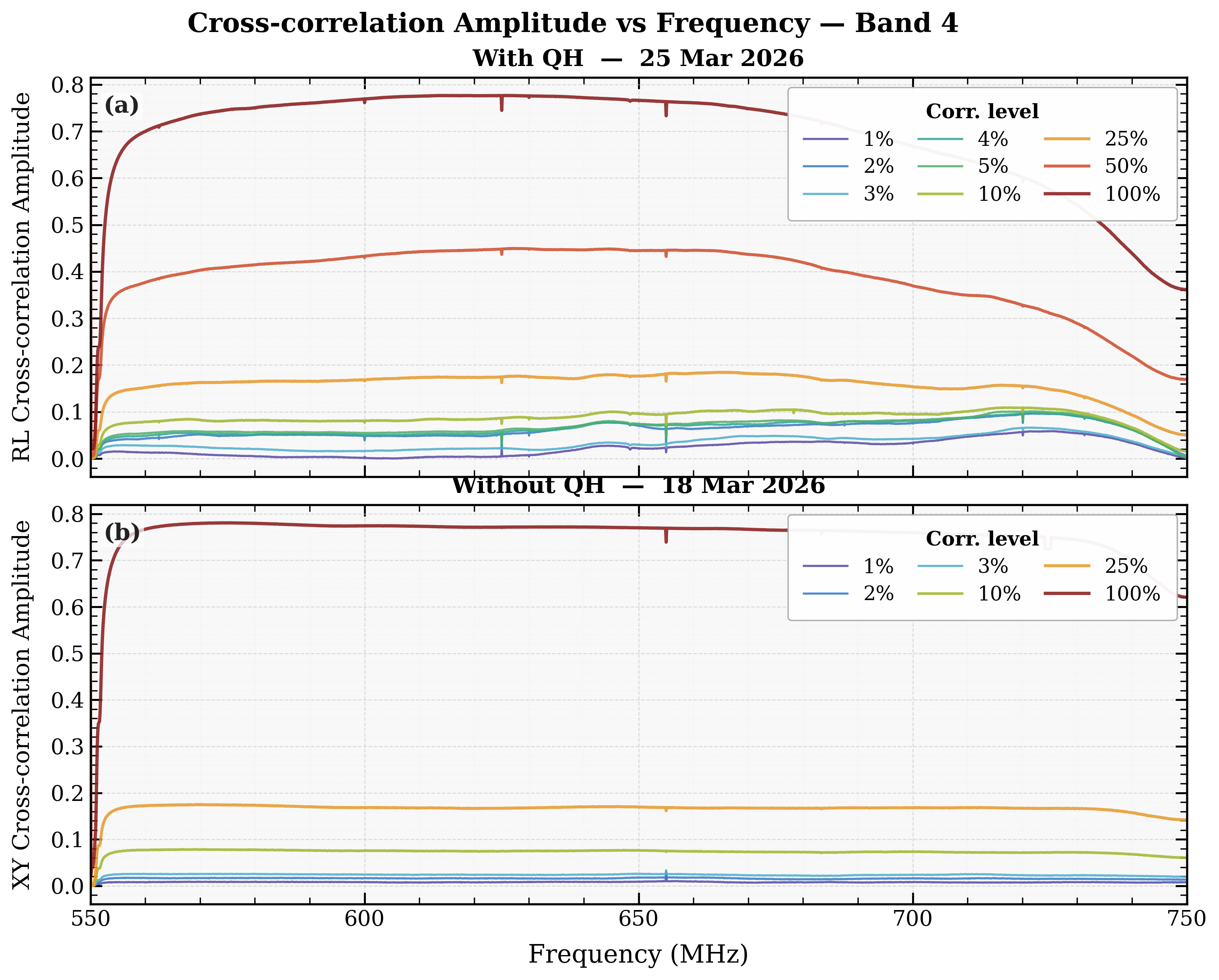}
\end{subfigure}
\hfill
\begin{subfigure}{0.48\textwidth}
    \centering
    \includegraphics[width=\linewidth]{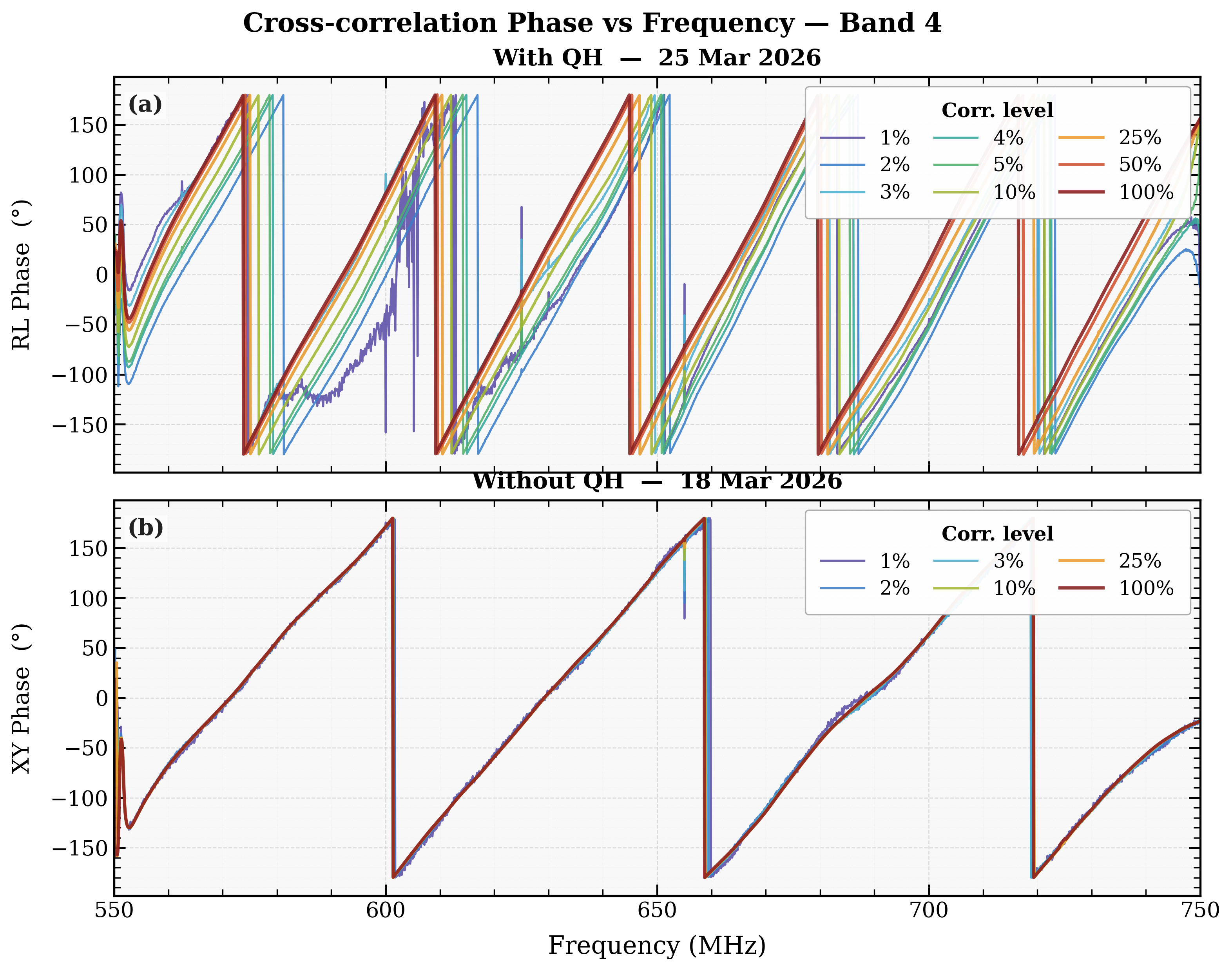}
\end{subfigure}
\caption{Cross-correlation characteristics for different correlation levels  over the 550 to 750 MHz. The left panel illustrates the cross-correlation amplitude as a function of frequency, indicating variations in signal magnitude for the two correlations. The right presents the corresponding phase response, showing the phase evolution. In every case, the top is the response of the system with the QH in place and the bottom is when the QH is bypassed.}
\label{sanjeet_noise2}
\end{figure*}

The results (Figure \ref{sanjeet_noise2}) show that without the QH, the system maintains stable amplitudes across all correlation strengths. With the QH, the system is more stable than in previous measurements but still exhibits greater amplitude oscillations compared to the without-QH configuration. Cross-phase analysis further highlights the differences. The with-QH system displays nonlinear, variable phase slopes at low correlations, which become linear at higher correlations. In contrast, the without-QH system behaves like an ideal radio interferometer frontend, with all cross-phases aligned. This demonstrates that the without-QH system can be calibrated reliably, whereas the with-QH system requires careful matching of the linear polarization strength between the calibrator and the target source.

\section{Discussion and Conclusions}\label{sec:conclusions}

The uGMRT is currently one of only two instruments (along with MeerKAT) capable of studying the universe at sub-GHz frequencies with sensitivities at the level of tens of $\mu$Jy. Among these, the uGMRT offers the highest angular resolution at these frequencies. Even with the advent of the Square Kilometre Array \citep{2024eas..conf..496D}, the uGMRT will remain the most sensitive sub-GHz radio telescope in the Northern hemisphere. This capability is essential for probing the magnetized universe through Faraday rotation and depolarization studies, phenomena that are uniquely accessible at these wavelengths \citep{2015A&ARv..24....4B, 2005A&A...441.1217B}. However, realizing the full scientific potential of these observations demands a signal chain whose polarimetric response is stable and well-characterized.

Our investigation has demonstrated that the Quadrature Hybrid polariser in the uGMRT Band~4 frontend introduces a systematic, source-dependent instrumental response that fundamentally undermines the standard assumption of calibration transferability. This finding has broad implications, both for the interpretation of existing data and for the design of future observing strategies.

The source-dependent behaviour of the QH polariser was not previously reported in the existing studies. This is likely because other systematic effects, incorrect calibrator models, uncorrected handedness \citep{2020arXiv200408542D, https://arxiv.org/abs/2310.04335}, and ionospheric Faraday rotation, dominated the error budget, or because existing studies focused on highly polarized sources such as pulsars, for which the QH response is more stable (as demonstrated in Figs.~\ref{qh_nqh_raw_phase} and~\ref{sanjeet_noise2}). 

Linear polarization studies with the uGMRT at these frequencies have been quite limited. In imaging-mode measurements, 3C286 is typically used as the polarization calibrator, whose model has recently been refined and must be supplied correctly \citep{hugo2024absolute}. From our experiments, the fractional polarization can be accurately recovered for sources with linear polarization strengths of a few percent, but the polarization angle remains unreliable, especially for weakly polarized or depolarized sources with polarization fractions below $\sim$10\%, which covers the majority of imaging science targets. The source linear polarization is not known \emph{a priori}, and even if it were, it would be impractical to find a calibrator matching the target's polarization strength.

Time-series measurements follow a different path, typically using pulsars that exhibit high linear polarization. As we have shown from the noise source measurements (Figure~\ref{sanjeet_noise2}) and the DA240 data (Figure~\ref{qh_nqh_raw_phase}, last column), the QH produces a delay slope whose nonlinearity diminishes with increasing fractional polarization strength. At fractional polarizations exceeding $\sim$25\%, the delay slopes change less abruptly and can potentially be calibrated if both the source and the calibrator pulsar have comparably high degrees of linear polarization. Nevertheless, this imposes a restrictive requirement that may not be generally satisfied.

The conversion from circular to linear feeds by bypassing the QH involves practical tradeoffs that merit discussion. The historical preference for circular feeds at sub-GHz frequencies was motivated by simpler ionospheric corrections, first-order leakage behaviour, and the convenience of phase-only beamforming for pulsar observations (Section~\ref{crisis}). However, these advantages must now be weighed against the demonstrated instability of the QH polariser over wide bandwidths. In the linear feed configuration, the parallactic angle rotation couples the gain and leakage terms, requiring an iterative calibration approach \citep{1996A&AS..117..149S, 2011A&A...527A.106S}. Ionospheric Faraday rotation manifests as amplitude decorrelation rather than a simple phase rotation, requiring amplitude corrections for coherent beamforming. These additional calibration requirements are well understood and routinely handled by modern software packages such as CASA \citep{2022PASP..134k4501C}. The critical advantage of the linear feed configuration is that the instrumental response does not depend on the polarization state of the observed source, enabling reliable transfer of calibration solutions from any calibrator to any target.

An alternative to physically bypassing the QH would be to develop a calibration framework that models the source-dependent instrumental response. In principle, if the QH's transfer function can be characterized as a function of the input polarization state, one could iteratively solve for both the true source polarization and the instrumental response. Such an approach would require a detailed electromagnetic model of the QH polariser, validated against laboratory measurements, and its incorporation into a direction-dependent calibration framework \citep{2011A&A...527A.106S}. While conceptually feasible, this strategy would impose significant computational overhead and introduce additional free parameters whose degeneracies with astrophysical signals (e.g., Faraday rotation, depolarization) could be difficult to break. We therefore consider the QH bypass to be the more practical and robust solution for the foreseeable future. In the case where the software pipeline requires circular-basis data, one can convert the calibrated linear-feed visibilities to a circular basis \citep{Perley2024}, provided the linear data are properly amplitude and phase balanced beforehand, and use it for subsequent rounds of self-calibration.

We can attempt to explain the fractional polarisation-dependent response of the QH analytically within the Jones matrix framework established in Section~\ref{sec:sigchain}. For a source with fractional polarisation $f_p$, the source coherency matrix decomposes 
exactly into a fully polarised component and an unpolarised component weighted by $f_p$ (Appendix~\ref{app:derivation}). Propagating this decomposition through the full system Jones matrix $\mathbf{M}(f) = \mathbf{M}_{\rm QH}(f)\,\mathbf{J}(f)$, the cross-correlation between the synthesised circular outputs L and R 
separates into two terms

\begin{equation}
    \phi_{\rm obs}(f,\,f_p) = \arg\!\left[f_p\,\mathcal{S}_{\rm direct}(f) 
    + (1-f_p)\,\mathcal{S}_{\rm leak}(f)\right]
    \label{eq:crossphase_main}
\end{equation}

\noindent where $\mathcal{S}_{\rm direct}$ carries the cross-coherency of the direct polarised signal path and $\mathcal{S}_{\rm leak}$ carries the cross-coherency of the leakage path. In the ideal case $\mathcal{S}_{\rm 
leak}$ vanishes identically and the cross-phase is independent of $f_p$.

In the real system, however, there exists a frequency-dependent differential 
gain between the two feed channels, $|G_x(f)| \neq |G_y(f)|$, which was 
established independently at the $\sim$10\% level from the bandpass 
solutions of the linear feed data and is not constant across the band. This 
frequency-dependent gain mismatch directly populates $\mathcal{S}_{\rm 
leak}$ with a non-zero contribution whose phase varies with frequency, as 
do the non-zero feed leakage terms $d_x(f)$, $d_y(f)$ measured in 
Section~\ref{sec:results}. Since $\mathcal{S}_{\rm direct}$ and 
$\mathcal{S}_{\rm leak}$ are complex quantities pointing in different 
directions in the complex plane, the observed cross-phase rotates 
continuously with $f_p$ as the weighted vector sum of these two phasors 
shifts direction. At high $f_p$ the direct signal term dominates and the 
cross-phase is stable and well-behaved. As $f_p$ decreases the leakage 
term progressively takes over and the cross-phase diverges, driven by the 
frequency-dependent gain mismatch and feed leakage. This is precisely the 
behaviour observed in Figures~\ref{qh_nqh_raw_phase} 
and~\ref{sanjeet_noise2}, where the cross-phase stabilises above 
$f_p \sim 50\%$ and becomes increasingly unstable at lower correlation 
levels.

Designing a better QH is also a possibility and out of scope of this investigation, but, \citet{Srikanth2001} explicitly caution that maintaining amplitude and phase balance across wide bandwidths is extremely demanding for these QH devices. Even a carefully optimised six-branch waveguide hybrid achieves amplitude imbalance $\leq 1$\,dB and phase imbalance $\leq 1^\circ$ only over a bandwidth ratio of $\sim$1.43:1, and tighter balance of 
$\leq 0.5$\,dB only over 1.29:1. To achieve even this, three separate hybrid designs with overlapping bands were required rather than a single broadband solution, underscoring how sensitive the balance is to frequency. 
The uGMRT Band~4 spans 550--750\,MHz, a ratio of 1.36:1, placing it 
directly within the same bandwidth regime. The degradation of amplitude and phase balance of the QH 
across the band therefore compounds the existing frequency-dependent gain 
mismatch of the feeds, and together these effects produce the systematic 
fractional polarisation-dependent cross-phase instability reported here. To our knowledge this effect has not previously been reported for any instrument employing a quadrature hybrid feed. Modern instruments such as the EVLA also employ quadrature hybrids in their signal chains, yet no published results demonstrate a fractional polarisation-dependent cross-phase response of the kind reported here. This is likely because the EVLA operates with physical sub-banding, where each band is divided into narrower sub-bands that individually satisfy the bandwidth constraints identified by \citet{Srikanth2001} more comfortably, alongside better gain matching 
between the two feed outputs achieved through tighter manufacturing tolerances and cryogenic cooling of the frontend electronics, which reduces both thermal noise and gain fluctuations across the band. Since the 
fractional polarisation-dependent response is driven by the frequency-dependent gain mismatch $|G_x(f)| \neq |G_y(f)|$ and the phase error $\delta(f)$ of the hybrid, any system that reduces these two quantities, whether through a narrower instantaneous bandwidth, better gain matching, or a more thermally stable frontend, will exhibit a weaker and more stable cross-phase response.

The lessons from this study also extend beyond the uGMRT. Many current and upcoming radio telescopes employ wideband receivers at sub-GHz frequencies. Our findings underscore the importance of end-to-end verification of polarimetric response across the full bandwidth, particularly when wideband polarisers are employed. The diagnostic methodology presented here provides a template for such verification at other facilities.

The key findings of this study are as follows:

\begin{enumerate}
    \item The QH polariser produces a cross-hand phase response that varies with the fractional polarization of the input signal. This source-dependent behaviour means that calibration solutions derived from one source cannot be reliably applied to another source with different polarization properties. In the with-QH uGMRT Band~4 system, accurate polarization calibration is only achievable when the calibrator and target have similar fractional polarizations, preferably exceeding 25\%.

    \item The QH introduces substantial instrumental leakage (10--15\%) with strong spectral modulations and significant antenna-to-antenna variations. With the QH bypassed, the leakage originates solely from the feed and is reduced to 2--5\%, with a flat spectral response and consistent behaviour across antennas.

    \item With the standard QH system, residual leakage after calibration is approximately 0.5\%, and the polarization angle spectrum does not reflect the true Faraday rotation of the source. The QH-bypassed system achieves residual leakage below 0.2\% and accurately recovers the expected polarization angle rotation. For DA~240, with a rotation measure of 3.3~rad~m$^{-2}$ \citep{2008A&A...489...69B}, the bypassed system correctly measures the $\sim 25^\circ$ rotation across the 550--750~MHz band.
\end{enumerate}

From a practical standpoint, our results demonstrate that reliable sub-GHz polarimetry with uGMRT Band~4 requires either bypassing the QH polariser or implementing a more sophisticated calibration strategy that accounts for the source-dependent instrumental response, assuming the response can be modelled. The QH-bypassed configuration converts the system to linear polarization feeds, which requires a modified calibration procedure but yields substantially improved polarimetric accuracy. For science cases that require accurate polarization angle measurements or rotation measure determinations, the linear feed configuration is strongly preferred. We note that the engineering modifications required to bypass the QH are straightforward.

This work emphasizes to look into instrumental problem more closely from both engineering and astronomical perspectives and pave the way for the future accurate polarimetric measurements with the uGMRT.

\section{Acknowledgments}
We thank Dipanjan Mitra for conceiving and designing the initial experiment
to debug the signal chain. He first identified the issue in pulsar
experiments, which subsequently guided the visibility-based investigation
presented here. His guidance was essential throughout the execution and
exploration of this work, and the entire author team expresses their sincere
gratitude to him. A.P. sincerely appreciates the discussions with him about calibration and test designing. We thank Atul Ganla and Ajith Kumar for designing the
correlated noise source used in the engineering tests. A.P. has been
involved in the sub-GHz polarization project for the past two years and
thanks Preshanth Jagannathan for the initial confirmation of the 3C286
polarization models from MeerKAT. A.P. is also grateful to Preshanth
Jagannathan, Shrikrishna Shekhar, and Sanjay Bhatnagar for their feedback
and discussions at various stages of this project. We thank Yashwant Gupta
and Sureshkumar S., and the other members of the Receiver Chain Issues group,
for their valuable feedback and continued support. A.P. acknowledges the
uGMRT technical staff who performed the QH bypass in the intense heat of
March at Khodad and appreciates being allowed to observe the process
firsthand from the cherry picker. A.P. personally thanks all the staff from
the Frontend, Backend, Servo, Electrical, and Operations groups who helped
facilitate these experiments. The authors acknowledge the use of ChatGPT (OpenAI) to assist in improving the clarity and readability of the manuscript. We thank the staff of the GMRT that made these observations possible. GMRT is run by the National Centre for Radio Astrophysics of the Tata Institute of Fundamental Research. This research has made use of NASA's Astrophysics Data System Bibliographic Services. This research has made use of the NASA/IPAC Extragalactic Database (NED),
which is operated by the Jet Propulsion Laboratory, California Institute of Technology,
under contract with the National Aeronautics and Space Administration.

\bibliographystyle{pasa}
\bibliography{cite}

@misc{2025ganla,
  author = {Ganla, Atul A.},
  title = {2-Channel Correlated Noise Source},
  year = {2025},
  note = {Internal Technical Report, Giant Metrewave Radio Telescope, NCRA-TIFR},
  url = {http://library.ncra.tifr.res.in:8080/jspui/handle/2301/573}
}

@misc{2025rai,
  author = {Rai, Sanjeet and Irappa, H. and Ganla, Atul},
  title = {A Report on Polarization Leakage Measurement},
  year = {2024},
  note = {Internal Technical Report, Giant Metrewave Radio Telescope, NCRA-TIFR},
  url = {http://library.ncra.tifr.res.in:8080/jspui/handle/2301/574}
}

@ARTICLE{1996A&AS..117..137H,
       author = {{Hamaker}, J.~P. and {Bregman}, J.~D. and {Sault}, R.~J.},
        title = "{Understanding radio polarimetry. I. Mathematical foundations.}",
      journal = {\aaps},
         year = 1996,
        month = may,
       volume = {117},
        pages = {137-147},
       adsurl = {https://ui.adsabs.harvard.edu/abs/1996A&AS..117..137H}
}

@ARTICLE{1996A&AS..117..149S,
       author = {{Sault}, R.~J. and {Hamaker}, J.~P. and {Bregman}, J.~D.},
        title = "{Understanding radio polarimetry. II. Instrumental calibration of an interferometer array.}",
      journal = {\aaps},
         year = 1996,
        month = may,
       volume = {117},
        pages = {149-159},
       adsurl = {https://ui.adsabs.harvard.edu/abs/1996A&AS..117..149S}
}

@ARTICLE{2011A&A...527A.106S,
       author = {{Smirnov}, O.~M.},
        title = "{Revisiting the radio interferometer measurement equation. I. A full-sky Jones formalism}",
      journal = {\aap},
         year = 2011,
        month = mar,
       volume = {527},
          eid = {A106},
        pages = {A106},
          doi = {10.1051/0004-6361/201016082},
archivePrefix = {arXiv},
       eprint = {1101.1764},
 primaryClass = {astro-ph.IM},
       adsurl = {https://ui.adsabs.harvard.edu/abs/2011A&A...527A.106S}
}

@ARTICLE{2017CSci..113..707G,
       author = {{Gupta}, Y. and {Ajithkumar}, B. and {Kale}, H.~S. and {Nayak}, S. and {Sabhapathy}, S. and {Sureshkumar}, S. and {Swami}, R.~V. and {Chengalur}, J.~N. and {Ghosh}, S.~K. and {Ishwara-Chandra}, C.~H. and {Joshi}, B.~C. and {Kanekar}, N. and {Lal}, D.~V. and {Roy}, S.},
        title = "{The upgraded GMRT: opening new windows on the radio Universe}",
      journal = {Current Science},
         year = 2017,
        month = aug,
       volume = {113},
       number = {4},
        pages = {707-714},
          doi = {10.18520/cs/v113/i04/707-714},
       adsurl = {https://ui.adsabs.harvard.edu/abs/2017CSci..113..707G}
}

@BOOK{1986isra.book.....T,
       author = {{Thompson}, A.~R. and {Moran}, J.~M. and {Swenson}, G.~W., Jr.},
        title = "{Interferometry and Synthesis in Radio Astronomy}",
         year = 2017,
      edition = {3rd},
    publisher = {Springer},
          doi = {10.1007/978-3-319-44431-4}
}

@ARTICLE{2022PASP..134k4501C,
       author = {{CASA Team} and {Bean}, B. and {Bhatnagar}, S. and {Castro}, S. and {Emonts}, B. and {Garcia}, E. and {Garwood}, R. and {Golap}, K. and {Harris}, P. and {Kepley}, A. and {Masters}, J. and {McNichols}, A. and {Mehringer}, D. and {Nakazato}, T. and {Ott}, J. and {Petry}, D. and {Rau}, U. and {Schiebel}, D. and {Tsutsumi}, T. and {van Bemmel}, I.},
        title = "{CASA, the Common Astronomy Software Applications for Radio Astronomy}",
      journal = {\pasp},
         year = 2022,
        month = nov,
       volume = {134},
       number = {1041},
          eid = {114501},
        pages = {114501},
          doi = {10.1088/1538-3873/ac9642}
}

@ARTICLE{1966MNRAS.133...67B,
       author = {{Burn}, B.~J.},
        title = "{On the depolarization of discrete radio sources by Faraday dispersion}",
      journal = {\mnras},
         year = 1966,
       volume = {133},
        pages = {67-83},
          doi = {10.1093/mnras/133.1.67}
}

@ARTICLE{2005A&A...441.1217B,
       author = {{Brentjens}, M.~A. and {de Bruyn}, A.~G.},
        title = "{Faraday rotation measure synthesis}",
      journal = {\aap},
         year = 2005,
        month = oct,
       volume = {441},
       number = {3},
        pages = {1217-1228},
          doi = {10.1051/0004-6361:20052990},
archivePrefix = {arXiv},
       eprint = {0507349},
 primaryClass = {astro-ph}
}

@ARTICLE{1998MNRAS.299..189S,
       author = {{Sokoloff}, D.~D. and {Bykov}, A.~A. and {Shukurov}, A. and {Berkhuijsen}, E.~M. and {Beck}, R. and {Poezd}, A.~D.},
        title = "{Depolarization and Faraday effects in galaxies}",
      journal = {\mnras},
         year = 1998,
       volume = {299},
        pages = {189-206},
          doi = {10.1046/j.1365-8711.1998.01782.x}
}

@ARTICLE{2015A&ARv..24....4B,
       author = {{Beck}, R.},
        title = "{Magnetic fields in spiral galaxies}",
      journal = {A\&ARv},
         year = 2015,
       volume = {24},
          eid = {4},
        pages = {4},
          doi = {10.1007/s00159-015-0084-4},
archivePrefix = {arXiv},
       eprint = {1509.04522},
 primaryClass = {astro-ph.GA}
}

@ARTICLE{2017JAI.....641011R,
       author = {{Reddy}, Suda Harshavardhan and {Kudale}, Sanjay and {Gokhale}, Upendra and {Halagalli}, Irappa and {Raskar}, Nilesh and {de}, Kishalay and {Gnanaraj}, Shelton and {Ajith Kumar}, B. and {Gupta}, Yashwant},
        title = "{A Wideband Digital Back-End for the Upgraded GMRT}",
      journal = {Journal of Astronomical Instrumentation},
         year = 2017,
        month = mar,
       volume = {6},
       number = {1},
          eid = {1641011-336},
        pages = {1641011-336},
          doi = {10.1142/S2251171716410117},
       adsurl = {https://ui.adsabs.harvard.edu/abs/2017JAI.....641011R}
}

@INPROCEEDINGS{2014ASInC..13..453S,
       author = {{Sureshkumar}, S.},
        title = "{Broadband feeds, frontend and fiber optic systems for the uGMRT}",
    booktitle = {Astronomical Society of India Conference Series},
         year = 2014,
       series = {Astronomical Society of India Conference Series},
       volume = {13},
        month = jan,
        pages = {453-456},
       adsurl = {https://ui.adsabs.harvard.edu/abs/2014ASInC..13..453S}
}

@techreport{hugo2024absolute,
  title={Absolute linear polarization angle calibration using planetary bodies for MeerKAT and JVLA at cm wavelengths},
  author={Hugo, B. and Perley, R.},
  year={2024},
  month={April},
  day={17},
  institution={South African Radio Astronomy Observatory and National Radio Astronomy Observatory},
  type={Technical Report},
  number={SSA-0004E-001, Revision C},
  url={https://doi.org/10.48479/bqk7-aw53},
  doi={10.48479/bqk7-aw53},
  note={Data release and calibration memo}
}

@INCOLLECTION{2021hai1.book..127R,
       author = {{Robishaw}, Timothy and {Heiles}, Carl},
        title = "{The Measurement of Polarization in Radio Astronomy}",
    booktitle = {The WSPC Handbook of Astronomical Instrumentation, Volume 1: Radio Astronomical Instrumentation},
         year = 2021,
       editor = {{Wolszczan}, Alex},
        pages = {127-158},
          doi = {10.1142/9789811203770_0006},
       adsurl = {https://ui.adsabs.harvard.edu/abs/2021hai1.book..127R}
}

@INPROCEEDINGS{2024eas..conf..496D,
       author = {{Diamond}, Philip},
        title = "{Square Kilometre Array Observatory}",
    booktitle = {EAS2024, European Astronomical Society Annual Meeting},
         year = 2024,
        month = jul,
          eid = {496},
        pages = {496},
       adsurl = {https://ui.adsabs.harvard.edu/abs/2024eas..conf..496D}
}

@ARTICLE{2020arXiv200408542D,
       author = {{Das}, Barnali and {Kudale}, Sanjay and {Chandra}, Poonam and {Bhattacharya}, Bhaswati and {Roy}, Jayanta and {Gupta}, Yashwant},
        title = "{The Polarization Convention of the uGMRT in Band 4}",
      journal = {arXiv e-prints},
         year = 2020,
        month = apr,
          eid = {arXiv:2004.08542},
        pages = {arXiv:2004.08542},
          doi = {10.48550/arXiv.2004.08542},
archivePrefix = {arXiv},
       eprint = {2004.08542},
 primaryClass = {astro-ph.IM},
       adsurl = {https://ui.adsabs.harvard.edu/abs/2020arXiv200408542D}
}

@article{https://arxiv.org/abs/2310.04335,
 adsurl = {https://ui.adsabs.harvard.edu/abs/2023arXiv231004335C},
 archiveprefix = {arXiv},
 author = {{Chandra}, Poonam and {Kumar}, S. Suresh and {Kudale}, Sanjay and {Kansabanik}, Devojyoti and {Das}, Barnali and {Kharb}, Preeti and {Silpa}, Sasikumar and {Sebastian}, Biny},
 doi = {10.48550/arXiv.2310.04335},
 eid = {arXiv:2310.04335},
 eprint = {2310.04335},
 journal = {arXiv e-prints},
 month = {October},
 pages = {arXiv:2310.04335},
 primaryclass = {astro-ph.IM},
 title = {{A detailed study of the polarisation convention of the Giant Metrewave Radio Telescope}},
 year = {2023}
}

@ARTICLE{2001A&A...375..344B,
       author = {{Bhatnagar}, S. and {Nityananda}, R.},
        title = "{Solving for closure errors due to polarization leakage in radio interferometry of unpolarized sources}",
      journal = {\aap},
         year = 2001,
        month = aug,
       volume = {375},
        pages = {344-350},
          doi = {10.1051/0004-6361:20010799},
archivePrefix = {arXiv},
       eprint = {astro-ph/0106348},
 primaryClass = {astro-ph},
       adsurl = {https://ui.adsabs.harvard.edu/abs/2001A&A...375..344B}
}

@ARTICLE{2008A&A...489...69B,
       author = {{Brentjens}, M.~A.},
        title = "{Deep Westerbork observations of Abell 2256 at 350 MHz}",
      journal = {\aap},
         year = 2008,
        month = oct,
       volume = {489},
       number = {1},
        pages = {69-83},
          doi = {10.1051/0004-6361:20077174},
archivePrefix = {arXiv},
       eprint = {0807.4467},
 primaryClass = {astro-ph},
       adsurl = {https://ui.adsabs.harvard.edu/abs/2008A&A...489...69B}
}

@ARTICLE{2026ApJS..283...82P,
       author = {{Perley}, Richard A. and {Butler}, Bryan J. and {Greisen}, Eric W. and {Hugo}, Benjamin V. and {Tremou}, Evangelia and {Willis}, A.~G.},
        title = "{Correcting Ionospheric Faraday Rotation for the VLA and MeerKAT}",
      journal = {\apjs},
         year = 2026,
        month = apr,
       volume = {283},
       number = {2},
          eid = {82},
        pages = {82},
          doi = {10.3847/1538-4365/ae503c},
archivePrefix = {arXiv},
       eprint = {2603.09001},
 primaryClass = {astro-ph.IM},
       adsurl = {https://ui.adsabs.harvard.edu/abs/2026ApJS..283...82P}
}

@techreport{Srikanth2001,
  author      = {Srikanth, S. and Kerr, A. R.},
  title       = {Waveguide Quadrature Hybrids for {ALMA} Receivers},
  institution = {National Radio Astronomy Observatory},
  year        = {2001},
  number      = {ALMA Memo 343},
  month       = jan,
  url         = {https://library.nrao.edu/public/memos/alma/main/memo343.pdf}
}

@techreport{Perley2024,
  author      = {Perley, R. and Greisen, E.},
  title       = {Post-Correlation Basis Conversion in {AIPS}},
  institution = {National Radio Astronomy Observatory},
  year        = {2024},
  number      = {EVLA Memo 229},
  month       = apr,
  url         = {https://library.nrao.edu/public/memos/evla/EVLAM_229.pdf}
}

\appendix

\section{Derivation of the Fractional Polarisation-Dependent 
Cross-Phase}\label{app:derivation}

\noindent We derive analytically the expression for the observed 
cross-phase between the synthesised circular outputs L and R as a function 
of fractional polarisation $f_p$, using the Jones matrix formalism 
introduced in Section~\ref{sec:sigchain}. For a source with total 
intensity $I$, fractional polarisation $f_p$ and electric vector position 
angle $\chi$, the Stokes parameters are $Q = f_p I\cos 2\chi$, 
$U = f_p I\sin 2\chi$ and $V = 0$ for a linearly polarised source. 
The source coherency matrix is therefore

\begin{equation}
    \mathbf{C}_{\rm src} = \frac{I}{2}
    \begin{pmatrix} 
        1 + f_p\cos 2\chi & f_p\sin 2\chi \\ 
        f_p\sin 2\chi     & 1 - f_p\cos 2\chi 
    \end{pmatrix}
\end{equation}

\noindent Applying the identities $1 + \cos 2\chi = 2\cos^2\chi$, 
$1 - \cos 2\chi = 2\sin^2\chi$ and $\sin 2\chi = 2\sin\chi\cos\chi$ to 
each element, this separates exactly into

\begin{equation}
    \mathbf{C}_{\rm src} = f_p\,\mathbf{C}_{\rm pol} + 
    (1-f_p)\,\mathbf{C}_{\rm noise}
    \label{eq:decomp}
\end{equation}

\noindent where $\mathbf{e} = (\cos\chi,\;\sin\chi)^T$ and

\begin{equation}
    \mathbf{C}_{\rm pol} = I\,\mathbf{e}\,\mathbf{e}^\dagger = 
    I\begin{pmatrix} 
        \cos^2\chi        & \sin\chi\cos\chi \\ 
        \sin\chi\cos\chi  & \sin^2\chi 
    \end{pmatrix}
\end{equation}

\begin{equation}
    \mathbf{C}_{\rm noise} = \frac{I}{2}
    \begin{pmatrix} 1 & 0 \\ 0 & 1 \end{pmatrix}
\end{equation}

\noindent $\mathbf{C}_{\rm pol}$ has rank 1 and zero determinant, 
which is the mathematical statement that the wave is fully coherent. 
$\mathbf{C}_{\rm noise}$ is proportional to the identity matrix, 
representing equal power in both channels with no cross-correlation, the 
mathematical statement of a fully incoherent wave. The decomposition in 
Equation~\ref{eq:decomp} is exact and carries no approximation.

\noindent The full system Jones matrix is 
$\mathbf{M}(f) = \mathbf{M}_{\rm QH}(f)\,\mathbf{J}(f)$, where the feed 
Jones matrix follows the notation of Equation~2

\begin{equation}
    \mathbf{J}(f) = 
    \begin{pmatrix} G_x(f) & d_x(f) \\ d_y(f) & G_y(f) \end{pmatrix}
\end{equation}

\noindent with $G_x(f)$ and $G_y(f)$ the complex gains of the X 
and Y channels and $d_x(f)$, $d_y(f)$ the complex leakage terms, all of 
which are in general frequency-dependent. Critically, the gain magnitudes 
$|G_x(f)|$ and $|G_y(f)|$ are not equal and their ratio varies across the 
band, producing a frequency-dependent power imbalance between the two 
channels that was established independently from the bandpass solutions of 
the linear feed data at the $\sim$10\% level. The quadrature hybrid Jones 
matrix for a branch-line coupler with phase error $\delta(f)$ is

\begin{equation}
    \mathbf{M}_{\rm QH}(f) = \frac{1}{\sqrt{2}}
    \begin{pmatrix} 
        1 &  i\,e^{i\delta(f)} \\ 
        1 & -i\,e^{i\delta(f)} 
    \end{pmatrix}
\end{equation}

\noindent where $\delta(f)$ encodes the departure of the hybrid 
from the ideal $90^\circ$ phase condition, which grows as the observing 
frequency departs from the design frequency $f_0$. The output coherency 
matrix is

\begin{equation}
    \mathbf{C}_{\rm out}(f) = \mathbf{M}(f)\,\mathbf{C}_{\rm src}\,
    \mathbf{M}^\dagger(f)
\end{equation}

\noindent Substituting Equation~\ref{eq:decomp} and using 
linearity of matrix multiplication

\begin{equation}
\begin{split}
    \mathbf{C}_{\rm out}(f) = \;& f_p\,\mathbf{M}(f)\,
    \mathbf{C}_{\rm pol}\,\mathbf{M}^\dagger(f) \\ 
    +\;& (1-f_p)\,\mathbf{M}(f)\,\mathbf{C}_{\rm noise}\,
    \mathbf{M}^\dagger(f)
\end{split}
\end{equation}

\noindent We now evaluate the two terms separately. Since 
$\mathbf{C}_{\rm pol} = I\,\mathbf{e}\,\mathbf{e}^\dagger$, the first 
term becomes $I\,(\mathbf{M}\mathbf{e})(\mathbf{M}\mathbf{e})^\dagger$. 
Computing $\mathbf{J}(f)\,\mathbf{e}$ explicitly

\begin{equation}
    \mathbf{J}(f)\,\mathbf{e} = 
    \begin{pmatrix} 
        G_x\cos\chi + d_x\sin\chi \\ 
        d_y\cos\chi + G_y\sin\chi 
    \end{pmatrix}
\end{equation}

\noindent The upper component is the direct response of the X 
channel to the incident polarised signal plus the leakage of the Y channel 
into X, and the lower component is the leakage of the X channel into Y 
plus the direct response of the Y channel. Applying $\mathbf{M}_{\rm QH}$ 
and denoting the two output components as $\alpha(f)$ and $\beta(f)$

\begin{equation}
\begin{split}
    \alpha(f) = \frac{1}{\sqrt{2}}\Big[&(G_x\cos\chi + d_x\sin\chi) \\ 
    +\;& i\,e^{i\delta(f)}(d_y\cos\chi + G_y\sin\chi)\Big]
\end{split}
\end{equation}

\begin{equation}
\begin{split}
    \beta(f) = \frac{1}{\sqrt{2}}\Big[&(G_x\cos\chi + d_x\sin\chi) \\ 
    -\;& i\,e^{i\delta(f)}(d_y\cos\chi + G_y\sin\chi)\Big]
\end{split}
\end{equation}

\noindent The off-diagonal element of 
$I(\mathbf{M}\mathbf{e})(\mathbf{M}\mathbf{e})^\dagger$ gives the direct 
signal cross-coherency

\begin{equation}
\begin{split}
    \mathcal{S}_{\rm direct}(f) 
    = \;& I\,\alpha(f)\,\beta^*(f) \\
    = \;& \frac{I}{2}\Big[(G_x\cos\chi + d_x\sin\chi)^2 \\
    +\;& e^{2i\delta(f)}(d_y\cos\chi + G_y\sin\chi)^2\Big]
\end{split}
\end{equation}

\noindent For the second term, since 
$\mathbf{C}_{\rm noise} = \frac{I}{2}\mathbf{I}$, we have 
$\frac{I}{2}\,\mathbf{M}(f)\,\mathbf{M}^\dagger(f)$. Computing 
$\mathbf{M}\,\mathbf{M}^\dagger = \mathbf{M}_{\rm QH}\,\mathbf{J}\,
\mathbf{J}^\dagger\,\mathbf{M}_{\rm QH}^\dagger$ and extracting the 
off-diagonal element gives the leakage cross-coherency

\begin{equation}
\begin{split}
    \mathcal{S}_{\rm leak}(f) = \frac{I}{4}\Big[&
    \left(|G_x|^2 + |d_x|^2\right) - 
    \left(|d_y|^2 + |G_y|^2\right) \\ 
    -\;& 2i\,e^{i\delta(f)}\,\mathrm{Re}\!\left[G_x d_y^* + 
    d_x G_y^*\right]\Big]
    \label{eq:sleak}
\end{split}
\end{equation}

\noindent Equation~\ref{eq:sleak} makes the physical origin of 
the leakage cross-coherency explicit. The first bracket, 
$(|G_x(f)|^2 + |d_x(f)|^2) - (|d_y(f)|^2 + |G_y(f)|^2)$, is the 
frequency-dependent power imbalance between the two channels. This term 
is non-zero whenever the total power in the X channel differs from that 
in the Y channel, and since $|G_x(f)|$ and $|G_y(f)|$ vary differently 
across the band, this imbalance has a non-trivial spectral structure even 
in the complete absence of any geometric feed leakage. The second bracket, 
$\mathrm{Re}[G_x(f)\,d_y^*(f) + d_x(f)\,G_y^*(f)]$, is non-zero 
whenever the leakage terms $d_x(f)$, $d_y(f)$ are non-zero, that is 
whenever there is any cross-coupling between the two polarisation channels 
in the feed. In the ideal case $d_x = d_y = 0$ and $|G_x(f)| = |G_y(f)|$ 
for all $f$, both brackets vanish identically and 
$\mathcal{S}_{\rm leak} = 0$ exactly. In the real system both conditions 
are violated across the observing band. The cross-correlation between L 
and R is the off-diagonal element of $\mathbf{C}_{\rm out}$

\begin{equation}
    \langle LR^*\rangle = f_p\,\mathcal{S}_{\rm direct}(f) + 
    (1-f_p)\,\mathcal{S}_{\rm leak}(f)
\end{equation}

\noindent and the observed cross-phase is therefore

\begin{equation}
    \phi_{\rm obs}(f,\,f_p) = \arg\!\left[f_p\,
    \mathcal{S}_{\rm direct}(f) + (1-f_p)\,\mathcal{S}_{\rm leak}(f)
    \right]
    \label{eq:crossphase}
\end{equation}

\noindent Since $\mathcal{S}_{\rm direct}$ and 
$\mathcal{S}_{\rm leak}$ are in general complex quantities with different 
phases, the argument of their weighted sum rotates continuously as $f_p$ 
changes, producing a cross-phase that depends systematically on the 
fractional polarisation of the observed source. The magnitude of this 
dependence is set by the angle between $\mathcal{S}_{\rm direct}$ and 
$\mathcal{S}_{\rm leak}$ in the complex plane, which is determined by the 
frequency-dependent gain mismatch $|G_x(f)|/|G_y(f)|$, the feed leakage 
amplitudes $|d_x(f)|$, $|d_y(f)|$, and the hybrid phase error $\delta(f)$. 
All three of these quantities are non-negligible in the uGMRT Band~4 
system across the 550--750\,MHz observing band, which is why the effect 
is clearly observable in the data presented in this paper.

\end{document}